\documentclass[aps,prf,twocolumn,groupedaddress,amsmath,amssymb,longbibliography,nofootinbib]{revtex4-1}
\usepackage{graphicx}  
\usepackage{dcolumn}   
\usepackage{bm}        
\usepackage{verbatim}   

\usepackage{braket}
\usepackage{pstricks}
\usepackage{epsfig}

\newcommand*{\Ek}{{\rm E}}
\newcommand*{\Ta}{{\rm Ta}}

\newcommand{\Ree}{{\rm Re}}
\newcommand{\Ro}{{\rm Ro}}
\newcommand*{\Ray}{{\rm Ra}}

\newcommand{\K}{{\rm K}}
\newcommand{\Bn}{{\rm B}_n}

\newcommand{\kv}{\ensuremath{\mathbf{k}}}
\newcommand{\rv}{\ensuremath{\mathbf{r}}}
\newcommand{\ve}{{\mathbf{v}}}
\newcommand{\lp}{\ensuremath{\left(}}
\newcommand{\rp}{\ensuremath{\right)}}

\begin{document}

\title{Thermal convection in rotating spherical shells: temperature-dependent internal\\ heat 
  generation using the example of triple-$\alpha$ burning in neutron stars}

\author{F. Garcia}
\affiliation{Department of Magnetohydrodynamics, Helmholtz-Zentrum Dresden-Rossendorf, Bautzner Landstra\ss e 400, D-01328 Dresden, Germany}
\affiliation{
Anton Pannekoek Institute for Astronomy, University of Amsterdam, Postbus 94249, 1090 GE Amsterdam, The Netherlands  
}

\author{F.~R.~N. Chambers}
\affiliation{
Anton Pannekoek Institute for Astronomy, University of Amsterdam, Postbus 94249, 1090 GE Amsterdam, The Netherlands    
}
\author{A.~L. Watts}
\affiliation{
Anton Pannekoek Institute for Astronomy, University of Amsterdam, Postbus 94249, 1090 GE Amsterdam, The Netherlands    
}

\date{\today}

\begin{abstract}
  We present an extensive study of Boussinesq thermal convection
  including a temperature-dependent internal heating source, based on
  numerical three-dimensional simulations. The temperature dependence
  mimics triple-$\alpha$ nuclear reactions and the fluid geometry is a
  rotating spherical shell. These are key ingredients for the study of
  convective accreting neutron star oceans. A dimensionless
  parameter $\Ray_n$, measuring the relevance of nuclear heating, is
  defined. We explore how flow characteristics change with increasing
  $\Ray_n$ and give an astrophysical motivation. The onset of
  convection is investigated with respect to this parameter and
  periodic, quasiperiodic, chaotic flows with coherent structures, and
  fully turbulent flows are exhibited as $\Ray_n$ is varied. Several
  regime transitions are identified and compared with previous results
  on differentially heated convection. Finally, we explore
  (tentatively) the potential applicability of our results to the
  evolution of thermonuclear bursts in accreting neutron star oceans.
\end{abstract}

\maketitle

\section{Introduction}

Convection is responsible for transporting angular momentum and for the
generation of magnetic fields in cosmic bodies - in particular
in the Earth's outer core~\cite{DoSo07}. Surface zonal patterns
observed on the gas giants (Jupiter and Saturn)~\cite{Chr02,HeAu07} and the
ice giants (Uranus and Neptune)~\cite{AHW07} are thought to be maintained by
convection within deeper layers. In the case of stars, the
differential rotation and meridional circulation observed in the
Sun~\cite{MBT06} and in main sequence stars~\cite{MCBB11} are modelled
using compressible convective models. The latter also seems to be quite
sufficient to explain the magnetic fields of isolated white
dwarfs~\cite{IGKL17}.

Convection, driven by nuclear burning, is also believed to be
important in accreting white dwarfs and neutron stars. For the former,
fully three-dimensional simulations in~\cite{JZNAB16} of the
convective dynamics establish the conditions for runaways
(thermonuclear explosions) in sub-Chandrasekhar-mass white dwarfs.
Neutron stars that accrete matter from a companion star build up a
low-density fluid layer of predominantly light elements (hydrogen and
helium) on top of the star's solid crust, forming a surface ocean
(see~\cite{ChHa08} for a discussion of the conditions for
solidification at the ocean-crust interface). Thermonuclear burning of
these elements as they settle in the ocean can be unstable, giving
rise to the phenomenon of Type I X-ray bursts (X-ray bursts are a
sudden increase in the X-ray luminosity of a source as radiation from
the runaway thermonuclear reactions escapes, see~\cite{Strohmayer06}
for a general review, and should not be confused with the convective
bursts studied in the planetary literature~\cite{Cho00,HeAu12}.) The
energy produced in the thermonuclear runaways cannot be dissipated by
radiative transfer, and then convection sets in. In the case of
neutron stars, surface patterns known as burst oscillations (for a
review see \cite{Wat12}) are observed to develop frequently during
these thermonuclear explosions, motivating an interest in convective
patterns~\cite{Wat12,Malone14,Keek17}.  Burst oscillations appear as
modulation of the X-ray luminosity in the aftermath of a Type I X-ray
burst. Possible explanations of this phenomenon involve flame
spreading~\cite{Spitkovsky02,Cavecchi13,Cavecchi15,Cavecchi16,Mahmoodifar16}
or global modes of oscillation of the ocean \cite{Hey04,PiBi05} (see
also~\cite{CWCGK18} in the case of superbursts) but we are still far
from having a complete understanding.

For the above-mentioned reasons many numerical, analytic and
experimental studies are devoted to this field. Good reviews can be
found in the literature, see for instance~\cite{Jon07} on convection
in the Earth's outer core,~\cite{Jon11} on planetary dynamos
or~\cite{Ols11} which focuses on experiments in rotating spherical
geometry. In the case of convective stellar interiors the
review~\cite{Gil00} gives a fluid dynamics perspective, focusing on
the effect of rotation in solar convection.  Finally, the recent
review of~\cite{ArMe16} describes the state of the art of stellar
simulations, covering a wide range of astrophysical applications.

Current astrophysical hydrodynamical numerical codes, incorporating
nuclear burning physics, are set up in square
geometries~\cite{Malone11,Malone14,ZMNAB15,Cavecchi13,Cavecchi15,Cavecchi16}
and are thus local in nature. For this reason the study of patterns of
convection in a rotating spherical geometry (global patterns)
generated by nuclear burning, as proposed in this paper, is novel and
of importance. The study is restricted to Boussinesq convection and
does not incorporate compositional gradients, which may be a
substantial simplification in the astrophysical context. However, this
approximation makes sense when the focus is to study basic
hydrodynamical mechanisms in tridimensional domains, as is commonly
adopted in the context of planetary atmospheres (see for instance
~\cite{HeAu07}). This knowledge will provide a starting point for
further global studies incorporating more complex physics, opening the
way for a deeper understanding of stellar processes.

Boussinesq and anelastic thermal convection in rotating spherical
shells with differential heating mechanisms have been studied in
considerable detail over the past
decades~\cite{Mac77,SuOl02,GGZ07,DHW16} (among many others), but in
the context of planetary cores and dynamos. The secular cooling of
planetary cores is modelled by internal buoyancy sources and the heat
flux is assumed to be nonuniform in the lateral direction of the outer
boundary, to mimic the thermal structure of the lower rocky
mantle. The physical characteristics of accreting neutron stars are
however quite different~\cite{Wat12}. The internal heat released in
thermonuclear burning reactions is strongly dependent on temperature:
helium flashes, which we consider in this paper, are caused by the
extremely temperature-sensitive triple-alpha reaction~\cite{Cla84}. In
addition, the flow velocity boundary conditions are stress-free rather
than zero. This is important for the generation of zonal patterns, as
has been shown in studies of planetary
atmospheres~\cite{AuOl01,Chr02,HeAu07}.

Changing mechanical boundary conditions from nonslip to stress-free,
or decreasing the gap width of the shell (more realistic for
convection in atmospheres), results in a strong zonal wind generation
and quasi-geostrophic flow, even at supercritical
regimes~\cite{AuOl01}. On the giant planets (Jupiter and Saturn) the
strong equatorial zonal flow is positive (prograde,
eastward)~\cite{Chr02,HeAu07} while in ice giants (Uranus and Neptune)
it is negative (retrograde, westward)~\cite{AHW07}. This transition
from prograde to retrograde zonal flow was interpreted in~\cite{AHW07}
as a consequence of vigorous mixing leading to a progressive
domination of inertial forces with respect to Coriolis forces. The
prograde zonal flow was also found to be quite robust when considering
the effect of density stratification~\cite{JoKu09}.

The focus of this study is to investigate thermal convection driven by
nuclear burning heat sources, as occurs in the envelopes of accreting
neutron stars, by means of direct numerical simulations (DNS) in a
rotating spherical shell geometry.  The convective patterns and their
mixing properties, arising prior to ignition, are important for the
modelling of thermonuclear bursts~\cite{Wat12}. Convection tends to
mix the fuel and ashes~\cite{WHCHPRFSBW04}, altering nuclear
reactions. We chose a very similar set-up to that used for planetary
atmospheres~\cite{GWA13}, although with a different buoyancy driving
mechanism, allowing a careful comparison. We find that the flows
excited by heat released from nuclear reactions exhibit relevant
features typical of planetary atmospheres.

The paper is organized as follows. In \S\ \ref{sec:model} we introduce
the formulation of the problem, and the numerical method used to
obtain the solutions. In \S\ \ref{sec:res} the type of solutions with
increasing $\Ray_n$ are described and we analyse their physical
properties, flow patterns, force balance and time scales. Section
\S\ \ref{sec:astr} contains a discussion of the application to
accreting neutron star oceans, and finally \S\ \ref{sec:conc}
summarises the results obtained.

\section{Model}
\label{sec:model}

We consider Boussinesq convection of a homogeneous fluid of density
$\rho$, thermal diffusivity $\kappa$, thermal expansion coefficient
$\alpha$, and dynamic viscosity $\mu$. The fluid fills the gap between
two concentric spheres, rotating about an axis of symmetry with
constant angular velocity ${\bf \Omega}=\Omega {\kv}$, and it is
subject to radial gravitational field ${\bf g}=-\gamma \rv$ ($\gamma$
is constant and $\rv$ the position vector). In the Boussinesq
approximation $\kappa$, $\alpha$ and $\mu$ are considered constants,
and the simple equation of state $\rho=\rho_0(1-\alpha(T-T_0))$ is
assumed in just the gravitational term. In the other terms a reference
state $(\rho_0,T_0)$ is assumed (see for instance~\cite{Ped79}).

Previous studies have mainly considered two different thermal driving
mechanisms~\cite{Cha81,DSJJC04}. Convection may be driven by an
imposed temperature gradient on the boundaries and/or by a uniform
distribution of heat sources $q$. In contrast to this, the present
study considers a temperature-dependent heat souce $q$, and fixed
temperature at the boundaries. Heat generation that depends on
temperature (and also on density and nuclear species mass fraction) is
used to model thermonuclear reactions in stellar
interiors~\cite{Cla84}.

In the following, we start by describing the widely-used system of
equations without internal heat generation, and afterwards introduce
the new system including a temperature-dependent heat source term.

\subsection{The equations and the method}
\label{sec:eq}

In the absence of internal heat sources, and considering perfectly
conducting boundaries, $T(r_i)=T_i$ and $T(r_o)=T_o$ ($r_i$ and $r_o$
being the radius of the inner and outer boundary, respectively), the
purely conductive state, in which the fluid is at rest, is given by
$\ve=0$ and $T_c(r)=T_0+\eta d \Delta T(1-\eta)^{-2}r^{-1}$, where
$\ve$ is the velocity field, $\eta=r_i/r_o$ the aspect ratio,
$d=r_o-r_i$ the gap width, $\Delta T=T_i-T_o$ the temperature
difference, and $T_0=T_i-\Delta T(1-\eta)^{-1}$, a reference
temperature.

Following the same formulation as in~\cite{SiBu03} the mass, momentum
and energy equations are written in the rotating frame of reference
and in terms of the velocity field $\ve$ and the temperature
perturbation from the conduction state $\Theta=T-T_c$. With units
$d=r_{o}-r_{i}$ for the distance, $\nu^2/\gamma\alpha d^4$ for the
temperature, and $d^2/\nu$ for the time, the equations are
\begin{align}
&\nabla\cdot\ve=0,\label{eq:cont}\\
&\partial_t\ve+\ve\cdot\nabla\ve+2\Ta^{1/2}\kv\times\ve = 
-\nabla p^*+\nabla^2\ve+\Theta\rv,\label{eq:mom}\\
&\Pr\lp\partial_t \Theta+\ve\cdot\nabla \Theta\rp= \nabla^2
\Theta+\Ray \eta (1-\eta)^{-2}r^{-3} \rv\cdot\ve,  \label{eq:ener}
\end{align}
where $p^*$ is the dimensionless pressure containing all the potential
forces. The centrifugal force is neglected since $\Omega^2/\gamma \ll
1$ in stellar interiors. The system is governed by four
non-dimensional parameters, the aspect ratio $\eta=r_i/r_o$ and the
Rayleigh $\Ray$, Prandtl $\Pr$, and Taylor $\Ta$ numbers. These
numbers are defined by
\begin{equation}
  \Ray=\frac{\gamma\alpha\Delta T d^4}{\kappa\nu},\quad
  \Ta^{1/2}=\frac{\Omega d^2}{\nu},\quad
  \Pr=\frac{\nu}{\kappa}.
\label{eq:param}
\end{equation}

If in addition to the externally imposed temperature gradient we are
also interested in considering internal heat sources the energy
equation~(\ref{eq:ener}) becomes
\begin{align}
\Pr\lp\partial_t \Theta+\ve\cdot\nabla \Theta\rp&= \nabla^2
\Theta+\Ray \eta (1-\eta)^{-2}r^{-3}\rv\cdot\ve \nonumber \\
&+\frac{\gamma\alpha d^6}{\nu^2\kappa}\frac{q}{C_p},\label{eq:ener2}
\end{align}
where $q$ is the rate of internal heat generation per unit mass and
$C_p$ the specific heat at constant pressure.

Rather that considering uniform internal heat generation $q$, the
focus of this paper is to study the effect of considering a
temperature-dependence.  Where this originates in nuclear reactions,
there are different types of temperature-dependence, depending on the
specific nuclear reaction ~\cite{Cla84}. We choose as an illustrative
example heat generation coming from the helium-burning triple-$\alpha$
reaction, which is thought to play a major part in thermonuclear
bursts in neutron star oceans. Hydrogen is also present in the neutron
star ocean, and can also burn and contribute to heat generation but in
the interests of simplicity we neglect this for now. Note that other
choices are possible, and the same procedure could be applied, for
instance, to carbon burning in the modelling of superbursts.  For the
specific internal heat source that we consider here, the nuclear
burning contribution from the helium triple-$\alpha$ reaction
(see~\cite{Cla84}) is
\begin{align}
  q&= 5.3\times 10^{18} \rho_5^2 \lp \frac{Y}{T_9} \rp^3 e^{-4.4/T_9} ~\text{erg}~ \text{g}^{-1}~ \text{s}^{-1}.
\label{eq:hel_bur}
\end{align}
In this equation $Y$ is the mass fraction of He and $T_9=f_9 T$ and
$\rho_5=f_5\rho$ are adimensional with $f_9=10^{-9}$ K$^{-1}$ and
$f_5=10^{-5}$ g$^{-1}$ cm$^{3}$. With the scales chosen for the
variables the new energy equation (from Eq.~\ref{eq:ener2}) takes the
form
\begin{align}
\Pr\lp\partial_t \Theta+\ve\cdot\nabla \Theta\rp&= \nabla^2
\Theta+\Ray \eta (1-\eta)^{-2}r^{-3}\rv\cdot\ve+q_n,\\
q_n&= \Ray_n\frac{1}{(\Theta+T_c)^3}e^{-\Bn/(\Theta+T_c)},
\label{eq:temp_Ra_n}
\end{align}
where the two new adimensional parameters are
\begin{equation}
\Ray_n=\frac{5.3\times 10^{45} \rho_5^2 Y^3 \gamma^4\alpha^4 d^{18}}{\nu^8 \kappa C_p}, \quad \Bn=\frac{4.4\times 10^{9}\alpha\gamma d^4}{\nu^2},
\end{equation}
$\Ray_n$ being the control parameter and the main driver of burning
convection.

We are interested in studying changes in convection when $\Ray_n$ is
increased from zero, and the rest of parameters are kept fixed. The
variation of $\Ray_n$ could be interpreted physically as a change in
the helium mass fraction $Y$.  The helium mass fraction should
decrease significantly over the course of a burst as it burns to
carbon (see the spherically symmetric numerical calculations of a pure
helium flash model in~\cite{WHCHPRFSBW04}), and assuming that it is
not replenished from overlying layers as a result of convective mixing
~\cite{WBS06}.  Both ~\cite{WHCHPRFSBW04} and ~\cite{WBS06} also found
a radial expansion of the convective zone within the ocean during the
burst.  Following unstable helium ignition, convection sets in the
base of the accreted helium layer, radially expands outwards, decaying
once when the burning rate becomes sufficiently slow later in the
burst. Assuming that this expansion of the extent of the convective
zone is sufficiently slow compared to the convective time scale, one
could also associate the variation of $\Ray_n$ with the variation of
the width $d$ of the convective layer. However, and in contrast to
varying $Y$, varying $d$ affects not only $\Ray_n$ but also $\Ta$ and
$\Ray$ (see eq.~\ref{eq:param}). We believe this is not a serious
inconvenience because the stronger dependence is $\Ray_n\sim d^{18}$
(rather than $\Ta\sim d^4$ and $\Ray\sim d^4$).

When $\Ray_n>0$ there is a new conductive state ($\ve=0$) which is
different from
\begin{eqnarray}
  T_c(r)=\frac{\Ray}{\Pr}\lp \frac{T_0}{\Delta T}+\frac{\eta}{(1-\eta)^2}\frac{1}{r}\rp,
\label{eq:cond_sta}
\end{eqnarray}
corresponding to the purely differentially heated case
($\Ray_n=0$). Because of the nonlinear temperature dependence of the
nuclear heat generation (Eq.~\ref{eq:hel_bur}) we have been not able
to find an analytic solution for the conductive state. However, as
will shown later on, the burning conductive state is also found numerically to be spherically symmetric, but with quite different radial
dependence compared to the differential heating case.

Note that without internal heat sources the equation for the
temperature perturbation~(\ref{eq:ener}) does not depend on $T_0/\Delta
T$, as this is eliminated when computing $\nabla T$. This is not generally true
when considering temperature-dependent internal heat
sources, and thus $T_0/\Delta T$ should be estimated according to the
problem of interest. For accreting neutron star oceans burning pure
helium it is reasonable to consider $T_0/\Delta T\sim O(1)$
(see for instance ~\cite{PiBi05,ZCN14}).
  
The equations are discretized and integrated as described
in~\cite{GNGS10} and references therein.  The solenoidal velocity
field is expressed in terms of the toroidal and poloidal potentials and
together with the temperature perturbation are expanded in spherical
harmonics in the angular coordinates. In the radial direction a
collocation method on a Gauss--Lobatto mesh is used. The boundary
conditions for the velocity field are stress-free, and perfectly
conducting boundaries are assumed for the temperature. The code is
parallelized in the spectral and in the physical space using OpenMP
directives. We use optimized libraries (FFTW3~\cite{FrJo05}) for the
FFTs in $\varphi$ and matrix-matrix products (dgemm
GOTO~\cite{GoGe08}) for the Legendre transforms in $\theta$ when
computing the nonlinear terms.

For the time integration, high order implicit-explicit backward
differentiation formulas (IMEX--BDF)~\cite{GNGS10,GNS13} are
used. In the IMEX method we treat the nonlinear terms explicitly, in
order to avoid solving nonlinear equations at each time step. The
Coriolis term is treated fully implicitly to allow larger time
steps. The use of \textit{matrix-free} Krylov methods (GMRES in our
case) for the linear systems facilitates the implementation of a
suitable order and time stepsize control.

\subsection{Output Data}
\label{sec:data}

We now introduce the output data analysed in this study, extracted from the DNS. The data emerge from the time series
of a property $P=P(t)$. The time average $\overline{P}$ of the time
series, or its frequency spectrum, can be computed once the solution
has saturated (reach the statistically steady state). The property $P$
may be global (obtained by volume-averaging), semi-global in which a
spatial average in certain direction has been carried out, or a purely
local property, measured at a point $(r,\theta,\varphi)$ inside the
shell.

The volume-averaged kinetic energy density is defined as 
$K=\frac{1}{2}\langle |\mathbf{v}|^2 \rangle_{\mathcal V}$, i.e.
\begin{equation*}
K=\frac{1}{\mathcal V}\int_{\mathcal V} \frac{1}{2}(\mathbf{v}
\cdot \mathbf{v}) \;dv.
\label{eq:ener_dens}
\end{equation*}
The axisymmetric ($K_a$) and the non-axisymmetric ($K_{na}$) kinetic
energy densities are defined by modifying accordingly the velocity
field in the previous volume integral. They are based, respectively,
on either the $m=0$ or all the $m\neq 0$ modes of the spherical harmonic
expansion of the velocity potentials. Similarly, the kinetic energy
density $K_m$ ($K_l$) restricted to a given order $m$ (degree $l$), alternatively
toroidal (poloidal) $K_T$ ($K_P$) kinetic energy densities can be
computed. The loss of equatorial symmetry can be studied by
considering the kinetic energy density contained in the symmetric part
of the flow, denoted as ${\rm K_s}$. In some cases, a combination of
these kinetic energy definitions, such as the axisymmetric toroidal
component $K^T_a$, will be used. The order (degree) $m_{\text{max}}$
($l_{\text{max}}$) for which $K_m$ ($K_l$) is maximum can be used to
infer length scales of the system. Better estimation $n$ of the latter
is provided in~\cite{ChAu06}: it is defined as the mean spherical
harmonic degree in the kinetic energy spectrum.

The Rossby number $\Ro$, measuring the relative importance of inertial
and Coriolis forces, is defined in the standard fashion, $\Ro=\Ree\Ek$, with
$\Ree=2\sqrt{K}$ and $\Ek=\Ta^{-1/2}$ the Reynolds and Ekman numbers,
respectively. Different definitions of $\Ro$ arise when considering
different components of $K$. For identifying the force balance taking
place in different flow regimes the volume averaged nongradient part
of the Coriolis, $\mathcal{\nabla\times F}_C$, viscous,
$\mathcal{\nabla\times F}_V$, Archimedian (i.e buoyancy), $\mathcal{\nabla\times
  F}_A$, and inertial, $\mathcal{\nabla\times F}_I$ forces are
obtained. They are computed as the kinetic energy density, but with the
corresponding part of the curl of the momentum equation, instead of
using the velocity field. By taking the curl, the pressure gradient
disappears from the force balance~\cite{OrDo14b,GOD17}.

An important quantity in geophysical and astrophysical fluid dynamics
is the so-called zonal flow, i.e, the azimuthally averaged azimuthal
velocity $\langle v_{\varphi} \rangle$ which is generically a function
of $(r,\theta)$. Several points inside the shell have been selected to
monitor the zonal flow. They are defined by combining different radial
positions $r=r_i+0.15d$, $r=r_i+0.5d$ and $r=r_i+0.85d$ with different
colatitudes $\theta=\pi/8$, $\theta=\pi/4$ and $\theta=3\pi/8$
(which are evenly distributed between the north pole and the
equator). At the same grid of points and $\varphi=0$, a probe for the
temperature $T$ has been set.

\subsection{Validation of Results and Numerical Considerations}
\label{sec:val_res}

The differential heating version of the code has been successfully
tested in~\cite{GNGS10} using the corresponding benchmark data
of~\cite{CACDGGGHJKMSTTWZ01}. The modification of the code to cope
with temperature dependent burning heat internal source is
straightforward, with minor modifications. This is because only the
evaluation of the burning rate $q_n$ is performed at the physical mesh
of points, as this term is strongly nonlinear.

A certain degree of accuracy in the time integration is necessary to
capture the right dynamics. This is especially important in the
oscillatory regime, where different attractors may be reached depending
on the tolerances required for the local time integration errors. For
this reason a variable size and variable order (VSVO) high order (up
to five) method is used in this study (see~\cite{GNGS10} for
details). The tolerances are $10^{-8}$ for the study of the onset and
oscillatory solutions and $10^{-5}$ for obtaining chaotic as well as
turbulent attractors. This has been shown to be sufficient in~\cite{GBNS14} for
different supercritical physical regimes observed in differentially
heated convection.

For obtaining time-averaged properties, initial transients, which may
be large close to the onset, are discarded. The number of measurements
has been selected to be large enough that the results do not change
significantly if the length of the time series is halved. For the
frequency analysis Laskar's algorithm of fundamental
frequencies~\cite{Las93} is used, which allows an accurate
determination of the frequencies with larger Fourier amplitudes of the
spectrum.

Given an angular discretization, the system is usually believed to be
well resolved if the kinetic energy decays by two orders of magnitude
in its spectrum (see~\cite{ChAu06} for instance). This is satisfied
for all of the DNS presented in this study obtained with
$N_r\in[40,60]$ and $L\in[64,192]$, being $N_r$ and $L$ the number of
radial collocation points and spherical harmonic truncation
parameter. The spatial resolution was increased from time to time in
order to look into spatial discretization errors. A brief numerical
study of the effects of the truncation parameters is reported in
Tables~\ref{table:mesh_study1} and~\ref{table:mesh_study2}. In the
former, some time-averaged properties are listed versus the spatial
resolutions for two different solutions with parameters $\Ta=2\times
10^{5}$, $\Pr=1$, $\eta=0.6$, $\Ray=6.2\times 10^{3}$ and $T_i/\Delta
T=1.5$. By increasing the resolution, reasonable errors of around
$5\%$ are obtained. This is the maximum threshold that is allowed, and
thus solutions from $\Ray_n\ge 10^{21}$ are computed with $N_r=50$ and
$L=128$, while those from $\Ray_n\ge 10^{26}$ are computed with
$N_r=60$ and $L=192$.

The basic conductive state radial profile has large derivatives close
to the boundaries which are very sensitive to the number of radial
discretization points. This is shown in Table~\ref{table:mesh_study2}
where the temperature at three different radial positions in the
equatorial plane, volume-averaged kinetic density, and
surface-averaged radial derivative of the temperature at the outer
surface is evaluated at specific instant of certain time. Although the
errors for the temperature and kinetic energy density are small at
$1\%$, errors of $20\%$ are obtained for the temperature radial
derivative when increasing $N_r=40$ to $N_r=100$. This should be taken
into account if some proxy based on this quantity is going to be
analysed in the future.

\begin{table}[]
$$
\begin{array}{lccccccccccc}
\hline  
\Ray_n  & N_r & L  &\overline{Ro} &\overline{Ro_p} &\overline{K/K_{na}} & \overline{n} & \overline{n_p} \\  
\hline
10^{21} & 40  & 64  & 0.1751        & 0.1030        & 1.268             & 9.498         & 12.816  \\
10^{21} & 50  & 128 & 0.1706        & 0.1014        & 1.200             & 9.641         & 12.691  \\
10^{26} & 50  & 128 & 0.9239        & 0.6411        & 1.495             & 16.093        & 21.401  \\
10^{26} & 60  & 192 & 0.9283        & 0.6346        & 1.535             & 15.090        & 20.625   \\
\hline
\end{array}					  
$$
\vspace{-0.3cm}
\caption{Burning Rayleigh number $\Ray$, number of radial points
  $N_r$, spherical harmonic truncation parameter $L$, mean Rossby and
  poloidal Rossby numbers, $\overline{Ro}$ and $\overline{Ro_p}$, mean
  ratio of the total to the nonaxisymmetric KEDs,
  $\overline{K/K_{na}}$, and time averaged mean total and poloidal
  spherical harmonic degree $\overline{n}$ and $\overline{n_p}$. The
  parameters are $\Ta=2\times 10^{5}$, $\Pr=1$, $\eta=0.6$,
  $\Ray=6.2\times 10^{3}$ and $T_i/\Delta T=1.5$}
  \label{table:mesh_study1}
\end{table}

\begin{table}[ht] 
  \begin{center}
    \begin{tabular}{lcccccc}
\vspace{0.1cm}            
& $N_r$ & $T_1$        & $T_2$       & $T_3$       & $K$         & $\partial_r T(r_o)$  \\
\hline\\
& $100$ & $52575.287$ & $65007.475$ & $40778.845$ & $1061.1348$ & $-4506586.0$  \\
& $80$  & $52573.245$ & $65009.777$ & $40783.231$ & $1061.1145$ & $-4412121.1$  \\
& $60$  & $52570.995$ & $65013.672$ & $40792.241$ & $1061.1036$ & $-4164847.9$  \\
& $40$  & $52056.756$ & $65106.535$ & $40533.784$ & $1075.9509$ & $-3570169.6$  \\
\hline
    \end{tabular}
\caption{Some properties after time-stepping $t=0.4317$ viscous units
  of time. The temperatures $T_1$, $T_2$ and $T_3$ are evaluated at
  $\varphi=0$, $\theta=3\pi/8$ and $r_1=r_i+0.15d$, $r_2=r_i+0.5d$ and
  $r_3=r_i+0.85d$, respectively (where $d=r_o-r_i$), and $K$ is the
  volume-averaged kinetic energy. The largest errors lie in the radial
  derivative of the temperature at the outer surface, $\partial_r
  T(r_o)$. The parameters are $\Ta=2\times 10^{5}$, $\Pr=1$,
  $\eta=0.6$, $\Ray=6.2\times 10^{3}$, $T_i/\Delta T=1.5$, and
  $\Ray_n=10^{20}$. The spherical harmonic truncation parameter is
  $L=64$; with $L=128$ the results are nearly the same. }
  \label{table:mesh_study2}
  \end{center}
\end{table}

\section{Results}
\label{sec:res}

Several studies have pointed out that in stellar interiors, convection
is believed to occur in thin layers of fluids having low Prandtl and
large Taylor numbers (see~\cite{GCW18} and references
therein). Radiative diffusion dominating viscosity in the Sun's
interior translates into $\Pr<10^{-3}$~\cite{BNS00} and the high
degree of electron degeneracy gives rise to very low $\Pr<10^{-3}$ in
convective layers of isolated white dwarfs~\cite{IGKL17}. For
analogous reasons, accreting neutron stars have oceans with very low
$\Pr<10^{-3}$ as well~\cite{GCW18}. Modelling large $\Ta$ and low
$\Pr$ is numerically challenging. According to~\cite{GCW18} the effect
of decreasing $\Pr$ (or increasing $\Ta$) for the onset of convection
results in an increase of $\omega_c$, giving rise to very small time
scales. In addition $m_c$ is especially large in the case of thin
shells~\cite{AHA04,GCW18} and thus very high spatial resolutions are
required for the DNS.

For this study we will consider a moderate Taylor number $\Ta=2\times
10^{5}$, Prandtl number $\Pr=1$ and aspect ratio $\eta=0.6$. In this
regime the preferred mode of differentially heated convection
($\Ray_n=0$) is spiralling columnar (see for instance ~\cite{Zha92})
with critical Rayleigh number $\Ray_c=6.180125\times 10^3$, drifting
frequency $\omega_c=-23.29847$ and azimuthal wave number $m_c=8$. The
computational requirements for studying the associated
finite-amplitude convection problem are still reasonably
affordable. In addition, most studies of spherical rotating
convection are at $\Pr \sim O(1)$ (see for instance~\cite{GWA13})
making easier the comparison with previous results.

Because we are interested in studying convection driven by helium burning
heating sources, rather than by differential heating, we choose a
Rayleigh number close to the onset $\Ray=6.2\times 10^3$. The
corresponding nonburning ($\Ray_n=0$) solution is a weakly nonlinear
rotating wave (RW) (also called a Rossby wave) with $m=8$ and frequency
$\omega=-23.26037$. This type of solution arises when the
spherical symmetry of the basic state has been lost via Hopf
bifurcations~\cite{EZK92}. We have also performed some tests with
subcritical $\Ray$ (even negative, i.e, with a stabilizing temperature
gradient) to see if convection can be driven when $\Ray_n$ is
increased from zero. In all cases we have found convective
solutions for a sufficiently large $\Ray_n$. Our simulations are
 performed mainly with $\Bn=1$ and $T_0/\Delta T=1.5$, but other values
were considered to check the robustness of the results.

In Sec.~\ref{sec:regI} the onset of burning convection and the first
instabilities giving rise to periodic and oscillatory solutions are
studied. The final part~\ref{sec:mean_prop} focuses more on the highly
supercritical $\Ray_n$ regime, characterized by highly chaotic and
turbulent solutions, and on the description of the physical properties
and patterns of the flow.

\subsection{First instabilities and oscillatory triple alpha convection}
\label{sec:regI}

The intention of this section, rather than accurately and exhaustively
performing the linear stability analysis (as in~\cite{GCW18}), is to
provide a first estimate of the critical values of $\Ray_n^c$,
frequencies $\omega_c$ and azimuthal wave numbers $m_c$, and to describe
the patterns and types of weakly supercritical flows by means of DNS
of the fully nonlinear equations. This approach is quite common in previous
studies of this field (see for instance~\cite{SiBu03} for uniform
internal heating sources or~\cite{Chr02} for differential heating).

For fixed $\Ray$ we first obtain the corresponding differentially
heated nonburning solution ($\Ray_n=0$) and then, starting from this
initial condition, we obtain a sequence of solutions by increasing
$\Ray_n$ successively by one order of magnitude (keeping the rest
of parameters fixed). We look for the first $\Ray_n>0$ at which the solution
is different (by measuring some proxy) from the initial condition,
i.e, from the nonburning solution at $\Ray_n=0$. An example of this
procedure is shown in Fig.~\ref{fig:cond_Ran}(a), where the time series
of the volume averaged kinetic energy density $K$ is displayed. By
starting from the $m=8$ RW (differentially heated) corresponding to
$\Ray=6.2\times 10^{3}$ and $\Ray_n=0$, the burning solution at
$\Ray_n=10^{18}$ tends (after a long transient) to a RW but with
$m=9$. We will describe the differences between these two
solutions. Before doing so, however, we must describe the conductive state when burning heat
sources are included.

\subsubsection{Conductive state and onset of convection}
\label{sec:cond}

\begin{figure}
\includegraphics[scale=1.15]{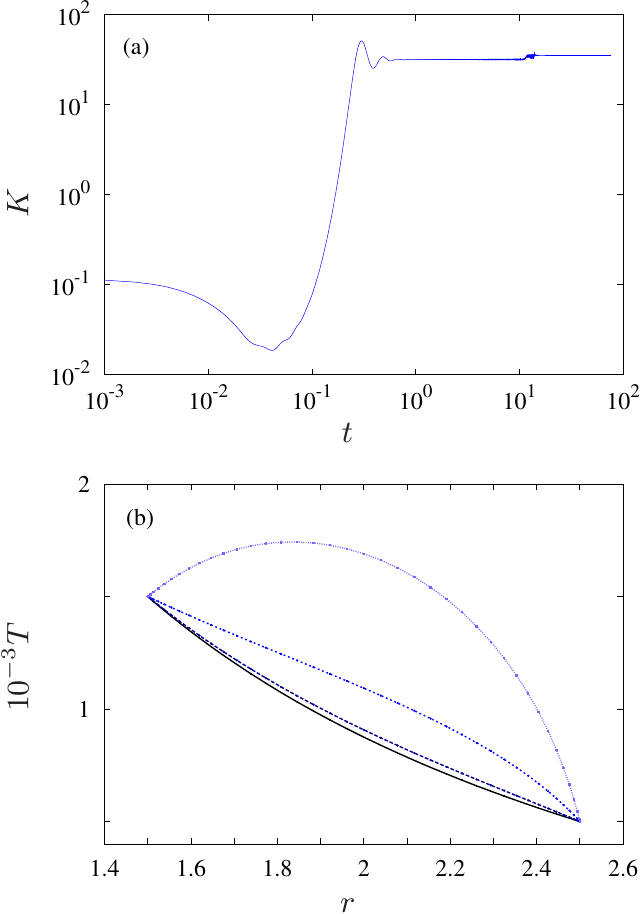}
  \caption{(a) Time series of the volume-averaged kinetic energy $K$
    at $\Ray=6.2\times 10^{3}$ and $\Ray_n=10^{18}$. The initial
    condition corresponds to $\Ray=6.2\times 10^{3}$ and
    $\Ray_n=0$. From a purely differentially heated convective flow
    and after a long transient ($t>20$) the burning flow attractor is
    reached (b) Conductive temperature $T$ versus radial coordinate
    $r$. The bottom curve corresponds to $\Ray_n=0$ with $T=T_c$
    given by Eq.~\ref{eq:cond_sta} (differentially heated conductive
    state). From bottom to top, the remaining curves correspond to
    $\Ray_n=10^{11}, 10^{12}, 10^{13}$, all with $\Ray=10^{3}$.}
\label{fig:cond_Ran}   
\end{figure}

Because $\Ray=-10^{3},10^0,10^1,10^2,10^3<\Ray_c$ and $\Ray_n=0$ the
conductive state given by Eq.~\ref{eq:cond_sta} is stable and thus any
velocity field perturbation decays to zero. By increasing $\Ray_n>0$
the burning conductive state, which is also spherically symmetric, is
obtained. Its radial profile is shown in Fig.~\ref{fig:cond_Ran}(b)
for $\Ray=10^3$ and $\Ray_n=10^{11}, 10^{12}, 10^{13}$. For
$\Ray_n<10^{11}$ the radial profile of the burning conductive state is
very similar to that of the nonburning case. This indicates that very
large $\Ray_n$ is required for convective onset. As will be shown in
Sec.~\ref{sec:astr}, large $\Ray_n$ are likely to occur in burning
stellar regions. From $\Ray>10^{12}$ the radial profile is
significantly different to that of $\Ray_n=0$, with an absolute
maximum of temperature close to the middle of the shell and very large
(modulus) derivatives close to the boundaries. This is similar to the
conductive profile ($\propto r^2$) obtained when constant internal
heating is considered~\cite{Cha81}.

\begin{figure*}
\includegraphics[scale=1.15]{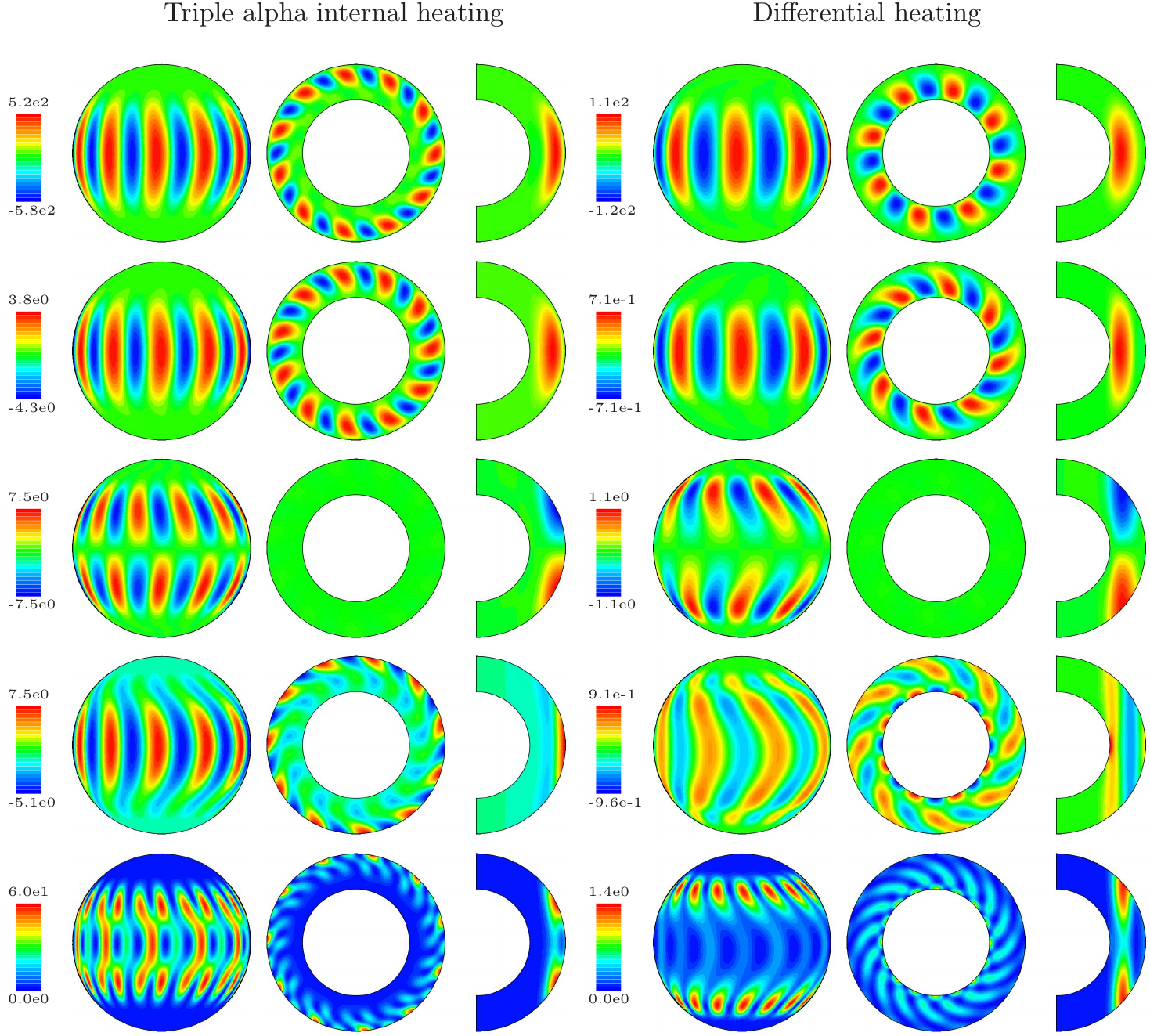}
\caption{First row: The left three plots are the spherical, equatorial
  and meridional cross-sections of the contour plots of the nonaxisymmetric
  component of the temperature $T$ at $\Ray=6.2\times 10^3$ and
  $\Ray_n=10^{17}$.  Right three plots: As the left, but for
  $\Ray=6.2\times 10^3$ and $\Ray_n=0$.  The second to fifth rows show $v_{r}$,
  $v_{\theta}$, $v_{\varphi}$ and $\ve^2/2$, from top to bottom.  All
  meridional cross-sections are selected at a relative maximum. The
  spherical cross-sections of $v_{\theta}$, $v_{\varphi}$ and $\ve^2/2$ are
  taken at the outer boundary. All of them cut a relative maximum
  except for $v_{\varphi}$ and $\Ray_n=0$ where the maximum is close
  to the inner boundary. The spherical cross-sections for $T$ and $v_r$ pass
  through the maximum, which is located inside the shell.}
\label{fig:cplt_RW}
\end{figure*}

For sufficiently large $\Ray_n$, and each different
$\Ray=-10^{3},10^0,10^1,10^2,10^3<\Ray_c$ explored, the spherically
symmetric burning conductive state becomes unstable to nonaxisymmetric
perturbations, giving rise to waves which drift azimuthally in the
prograde direction as in the case when $\Ray_n=0$. In all cases, the
 critical burning Rayleigh number required for convective onset is
of order $\Ray_n^c\sim 10^{17}$, with critical azimuthal wave numbers
$m_c\sim 10$ and critical drift frequencies $\omega_c\sim -50$. For
instance, at $\Ray=-10^3$ and $\Ray_n=10^{18}$ an $m=10$ RW with
$\omega=71.5$ is found, and at $\Ray=10^3$ and $\Ray_n =1.3\times
10^{17}$ the RW has azimuthal symmetry $m=11$ and $\omega=48.9$. These
values should not differ so much from the critical values, because at
$\Ray_n=10^{17}$ the conductive state is found to be stable. These results
suggest that the $\Ray_n^c$ required for the onset of burning convection
depends on the particular $\Ray$ chosen, and thus on the temperature
difference between the boundaries.

By increasing $\Ray_n$ up to $10^{12}$, and integrating the equations,
starting from the differentially heated nonburning initial condition
(RW, $m=8$) but also from random fields, the same solution (RW, $m=8$)
is obtained. Beyond $\Ray_n=10^{12}$ this Rossby wave is progressively
lost (decreasing the magnitude of the velocity field) and at
$\Ray_n\in[10^{14},10^{16}]$ the burning conductive state becomes
stable again.  The latter observation means that in convective systems
with internal heat sources, the onset of differentially heated
convection (measured by $\Ray_c$) depends on the internal heating rate
(measured by $\Ray_n$). A reasonable physical interpretation of this
restabilization is as follows: because $\Ray=6.2\times 10^3$ is very
close to the onset the heat flux is almost conductive, giving a
temperature profile similar to the solid black line in
Fig.~\ref{fig:cond_Ran}(b); by increasing $\Ray_n$ temperature
profiles become more parabolic-like with heat flux larger at the outer
boundary. In between, there exist some $\Ray_n$ for which heat flux is
rather uniform in the fluid layer (small dashed curve of
Fig.~\ref{fig:cond_Ran}(b)) making difficult to excite convective
motions.

At $\Ray_n=10^{17}$ neither the burning conductive state nor the
thermal $m=8$ RW are found. The solution, which is also a RW with $m=11$
and $\omega=55.1$, is quite different from the thermal wave found previously. At $\Ray_n
=10^{18}$ the same type of RW, with azimuthal symmetry $m=9$ and
$\omega=74.6$, is obtained after a long transient (see
Fig.~\ref{fig:cond_Ran}(a)). These waves, driven by triple alpha
heating, are of the same type as that seen at $\Ray<\Ray_c$
and $\Ray_n\gtrsim 10^{17}$ .

The comparison of the flow patterns between the triple alpha and
differential heating case is shown in Fig.~\ref{fig:cplt_RW}. The
latter displays the contour plots of the nonaxisymmetric component of
the temperature $T$, radial velocity $v_{r}$, colatitudinal velocity
$v_{\theta}$, azimuthal velocity $v_{\varphi}$ and kinetic energy
density $\ve^2/2$ from top to bottom row, respectively, in
spherical, meridional and equatorial cross-sections (see figure
caption). The left group of three cross-sections corresponds to the triple
alpha heating $m=11$ RW (at $\Ray=6.2\times 10^3$ and
$\Ray_n=10^{17}$) while the right group corresponds to the thermal
(Rossby) $m=8$ RW (at $\Ray=6.2\times 10^3$ and $\Ray_n=0$). Thermal
(Rossby) waves characteristic of $\Pr=1$ have been widely described
before for differential as well as internal heating
models~\cite{Zha92,SiBu03,GCW18} so we comment on these only briefly. These modes have been named spiralling columnar (SC)
because the flow is aligned with the rotation axis, forming convective
columns that are tangential (or nearly so) to the inner cylinder (see
right group of plots in Fig~\ref{fig:cplt_RW}). In contrast,
convection in triple alpha heating modes is attached to the outer
sphere and reaches lower latitudes including the equator, similar to
the equatorially attached EA modes characteristic of lower $\Pr<1$
which are more localised at equatorial
latitudes~\cite{Zha93,Zha94,GCW18}.  Triple alpha convective modes
also exhibit weak vertical dependence because of the Taylor-Proudman
constraint~\cite{Ped79}, given the moderate, but sufficiently large, Taylor
number $\Ta=2\times 10^5$.

\subsubsection{Oscillatory burning convection}
\label{sec:osc}

\begin{figure}
\includegraphics[scale=1.00]{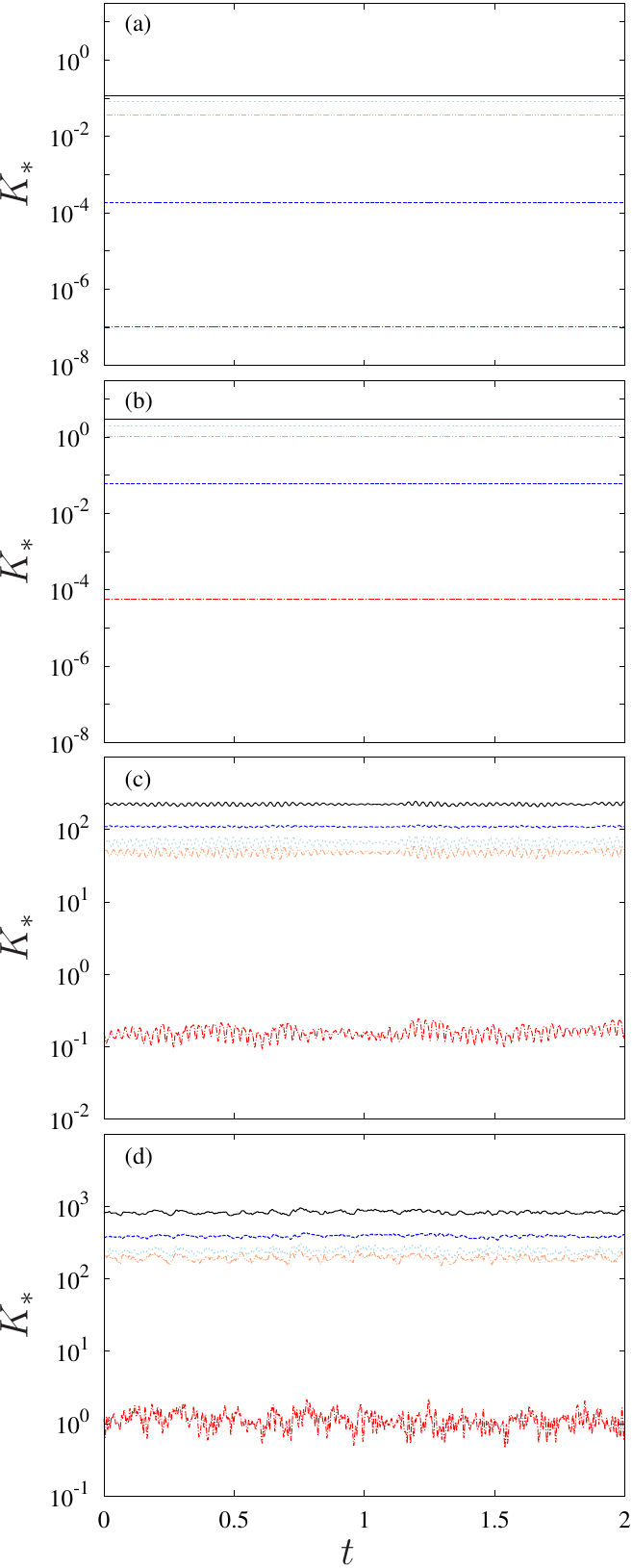}  
\caption{Time series of the kinetic energies $K$, $K_{a}^{T}$,
  $K_{na}^{T}$, $K_{a}^{P}$, and $K_{na}^{P}$ with solid, dashed (blue
  online), dotted (light-blue online), dashed-dotted (red online), and
  dashed-double-dotted (light-red online) line style,
  respectively. The parameters are $\Ray=6.2\times 10^{3}$ and (a)
  $\Ray_n=0$, (b) $\Ray_n=10^{17}$, (c) $\Ray_n=10^{19}$ and (c)
  $\Ray_n=10^{20}$.}
\label{fig:ener_t} 
\end{figure}

The time series of the volume-averaged kinetic energy density $K$, and
its different components $K_{a}^{T}$, $K_{na}^{T}$, $K_{a}^{P}$, and
$K_{na}^{P}$ displayed in Fig.~\ref{fig:ener_t}(a,b) correspond to the
thermal $m=8$ RW at $\Ray=6.2\times 10^3$ and the burning $m=11$ RW at
$\Ray=6.2\times 10^3$ and $\Ray_n=10^{17}$. They are constant because
azimuthally averaged properties do not change with time, as the flow
drifts in that direction. Both solutions have stronger toroidal
components and are almost nonaxisymmetric
$K_{a}^{P}<K_{a}^{T}<K_{na}^{P}<K_{na}^{T}$ because they are very
close to the respective onset. Increasing $\Ray_n$ results in a strong
relative increase of zonal motions measured by $K_{a}^{T}$, giving
$K_{a}^{P}<K_{a}^{T}\approx K_{na}^{P}<K_{na}^{T}$ at $\Ray_n=10^{18}$
(not shown in the figure) and $K_{a}^{P}< K_{na}^{P}\lesssim
K_{na}^{T}<K_{a}^{T}$ at $\Ray_n=10^{19}$ and $\Ray_n=10^{20}$ (see
Fig.~\ref{fig:ener_t}(c,d)). In all cases $K_{a}^{P}< 10^{-3}K$,
meaning that the axisymmetric component is almost purely toroidal.
The time series of Fig.~\ref{fig:ener_t}(c,d) exhibits oscillatory and
chaotic dependence with a strong axisymmetric and toroidal component
(zonal wind). All of these features of convection are typical with
stress-free boundary conditions, and have been reported before with
internal heating sources~\cite{SiBu03} as well as externally forced
temperature gradients~\cite{Chr02}. The description of their flow
patterns is delayed to Sec.~\ref{sec:cont_pl}, where turbulent
solutions will be also detailed.

\begin{figure}
\includegraphics[scale=1.00]{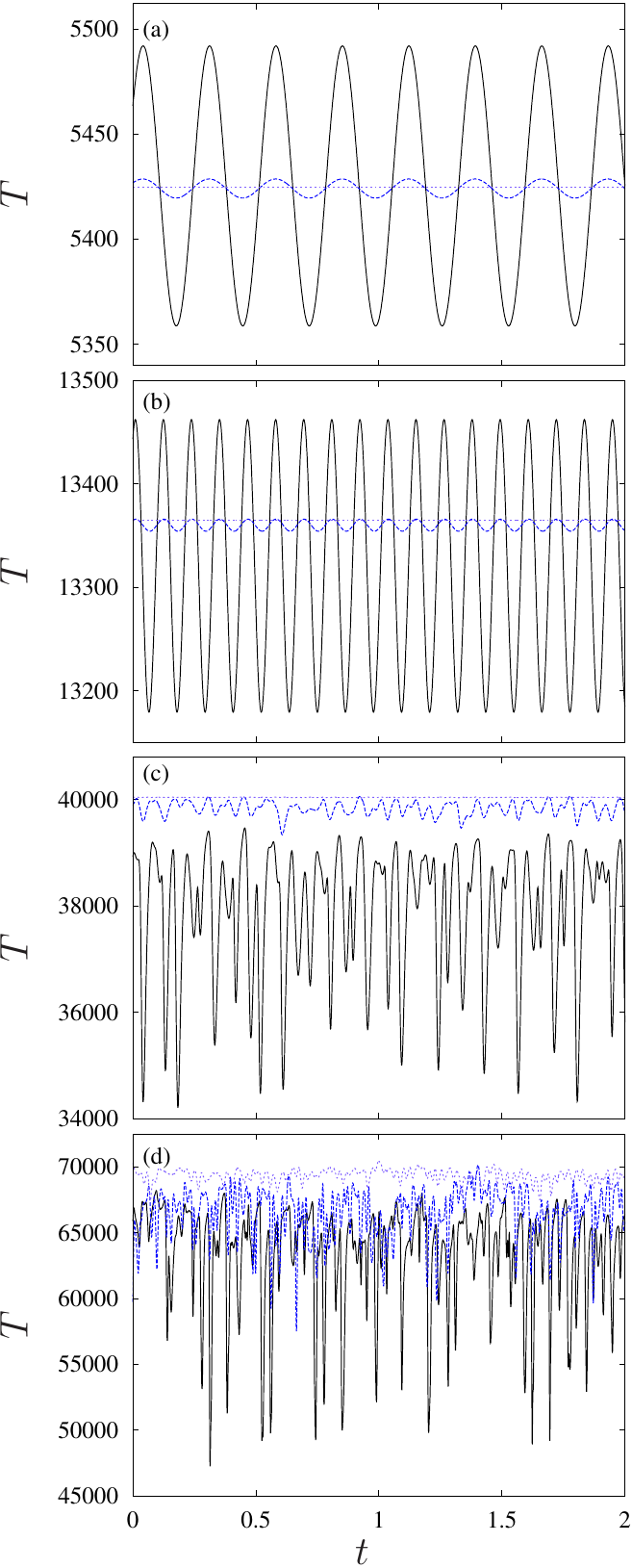}
\caption{Time series of the temperature at $r=r_i+0.5d$, $\varphi=0$
  and three different colatitudes $\theta=3\pi/8$, $\theta=\pi/4$ and
  $\theta=\pi/8$, with solid (black online), dashed (blue online),
  dotted (light-blue online) line style, respectively. The parameters
  are $\Ray=6.2\times 10^{3}$ and (a) $\Ray_n=0$, (b)
  $\Ray_n=10^{17}$, (c) $\Ray_n=10^{19}$ and (c) $\Ray_n=10^{20}$.}
\label{fig:temp_t}
\end{figure}

The time series of the temperature measured in the middle of the shell
($r=r_i+0.5d$) at $\varphi=0$ and three different colatitudes, evenly
distributed between the equatorial plane and the north pole
$\theta=\pi/8$, $\theta=\pi/4$ and $\theta=3\pi/8$ are displayed in
Fig.~\ref{fig:temp_t}(a,b,c,d) for the same solutions as
Fig.~\ref{fig:ener_t}(a,b,c,d), respectively. For the $m=8$ RW at
$\Ray_n=0$ and the $m=11$ RW at $\Ray_n=10^{17}$ the temperature is
periodic and the amplitude of the oscilations is around $3\%$ of the
mean (see Fig.~\ref{fig:temp_t}(a,b)) meaning that these solutions
represent small deviations from the conductive state. Both the amplitude of
the oscillations and the mean decrease with increasing $\theta$, and
the temperature in the polar regions ($\theta=\pi/8$) is almost
constant (i.e conductive). With increasing $\Ray_n$ the amplitude of
oscillations becomes larger (up to around $30\%$ at $\Ray_n=10^{20}$,
see Fig.~\ref{fig:temp_t}(d)) and the temperature in the polar regions
becomes more oscillatory. The strongest temperature
oscillations in our model of triple alpha burning convection are
located near the equatorial region, which may have observational
consequences. In addition, the time series of chaotic solutions shown
in Fig.~\ref{fig:temp_t}(c,d)) exhibits small rapid fluctuations,
coexisting with these large oscillations which have a longer associated
time scale. This longer time scale seems not to vary much with
$\Ray_n$ and is similar to the corresponding periodic oscilations
at the onset (compare Fig.~\ref{fig:temp_t}(b) with
Fig.~\ref{fig:temp_t}(d)). A deeper study of the time scales
associated with burning driven flows is considered in
Sec.~\ref{sec:flw_ts}.

\subsection{High $\Ray_n$ convection and mean properties}
\label{sec:mean_prop}

Time averaged global and local physical properties, such as volume
averaged kinetic energy densities or temperature at a point inside the
shell, as well as flow patterns, are described and interpreted in the
following subsection. In addition, some power laws derived from the equations of
motion will be compared to the numerical results, by assuming
certain force balances~\cite{GWA16,GOD17}. This has been shown to be a
successful tool for obtaining estimations of realistic
phenomena~\cite{ABNCM01,Chr02,GSN14,OrDo14} in the geodynamo context.
The analysis is restricted to $\Ray=6.2\times 10^3$, and comprises a
large sequence of solutions, including those studied in the previous
section, with increasing $\Ray_n$ up to $10^{27}$. This section is
focused on high $\Ray_n$ flows and on detailing the associated
physical regimes. By increasing $\Ray_n$, turbulent convection develops
after different transitions from the laminar regime. We have found
that burning driven convection shares relevant flow features with the
type of convection described for planetary
atmospheres~\cite{HeAu07,AHW07,GWA13} modelled with an externally
forced temperature gradient. In particular, we find the same regimes described
in~\cite{GWA13} (see the summary in their conclusions) for both
Boussinesq and anelastic approximations.

Table~\ref{table:prop_mean_ener} contains data that give a preliminary
description of the route to turbulence occuring in triple alpha
burning convection. The Rossby number, and its poloidal component
measuring vigour of convective flow~\cite{Chr02,GSN14}, increase
monotonically.  For the small $\Ray_n$, $Ro\ll 1$ and $Ro_p\approx
Ro/2$ indicates that Coriolis forces are important, and the flow is
moderately convective. However, by increasing nuclear burning rate
$Ro_p\lesssim\Ro\sim 1$, thus Coriolis forces play secondary role and
the flow is strongly convective. The ratio $K/K_{na}$ helps to
identify these different flow regimes~\cite{GWA13}, decreasing from
$\Ray_n=10^{23}$ for the largest $\Ray_n$, and provides useful
information. For instance, $\K/\K_{na}$ is also a maximum in the
vicinity of $\Ray_n=10^{19}$, the regime of oscillatory and chaotic
(but coherent) burning convection studied previously. The spherical
harmonic order $m_{\text{max}}$, in which $\max K_m$ is reached, or
equivalently $l_{\text{max}}$ for the spherical harmonic degree, are
often used for estimating length scales present in the system. More
accurate estimations could be obtained with with
$(\overline{n})^{-1}$~\cite{ChAu06} (or $(\overline{n_p})^{-1}$ for
convective length scales). This table points to so-called Rossby-wave
turbulence, i.e the low wave number route to chaos
(see~\cite{RuTa71,Eck81} for examples of routes) in which the energy
is contained at successively lower wave numbers when increasing the
forcing (inverse cascade, because of the shearing produced by
differential rotation~\cite{ScCa05}). Regarding the regimes in the
extremes (rotating waves at $\Ray_n=10^{17},10^{18}$ or the most
turbulent solutions from $\Ray_n\ge10^{25}$), large length scales that
do not change substantially are found, and the convective flow
develops smaller scale structures.

\begin{table}
$$
\begin{array}{lccccccccccc}
\hline  
\Ray_n  &\overline{Ro} &\overline{Ro_p} &\overline{K/K_{na}} & m_{\text{max}} & l_{\text{max}}  & \overline{n} & \overline{n_p} \\  
\hline
0       & 0.001        & 0.0006         & 1.002             & 8,0,16       & 9,11,3       & 9.44         & 8.31  \\
10^{17}  & 0.006        & 0.003          & 1.02              & 11,0,22      & 12,14,3      & 11.70        & 11.14  \\
10^{18}  & 0.019        & 0.010          & 1.32              & 9,0,18       & 10,3,1       & 8.46         & 9.62  \\
10^{19}  & 0.047        & 0.022          & 1.97              & 0,8,16       & 3,1,9        & 8.47         & 14.67  \\
10^{20}  & 0.091        & 0.044          & 1.89              & 0,5,10       & 3,1,5        & 8.09         & 13.63  \\
10^{21}  & 0.17         & 0.101          & 1.20              & 1,0,3        & 3,9,12       & 9.64         & 12.69  \\
10^{22}  & 0.32         & 0.16           & 2.53              & 0,2,5        & 3,1,5        & 7.06         & 14.92  \\
10^{23}  & 0.50         & 0.25           & 2.88              & 0,2,4        & 3,1,7        & 7.06         & 15.81  \\
10^{24}  & 0.68         & 0.35           & 2.74              & 0,2          & 3,1,5        & 8.19         & 17.41  \\
10^{25}  & 0.81         & 0.48           & 2.06              & 0,2          & 3,1,5        & 11.28        & 19.52  \\
10^{26}  & 0.92         & 0.64           & 1.50              & 0,5          & 3,1,5        & 16.09        & 21.40  \\
10^{27}  & 1.10         & 0.84           & 1.20              & 0,5          & 1,3,6        & 21.23        & 23.07  \\
\hline
\end{array}					  
$$
\caption{Burning Rayleigh number $\Ray_n$, mean Rossby and poloidal
  Rossby numbers, $\overline{Ro}$ and $\overline{Ro_p}$, mean ratio of
  the total to the nonaxisymmetric KEDs, $\overline{K/K_{na}}$,
  leading azimuthal wave numbers having a relative maximum of
  $\overline{K_m}$ (the KED considering a single wave number $m$),
  leading spherical harmonic degrees having a relative maximum of
  $\overline{K_l}$ and time averaged mean total and poloidal spherical
  harmonic degree $\overline{n}$ and $\overline{n_p}$.}
\label{table:prop_mean_ener}
\end{table}

\subsubsection{Time-averaged Properties}
\label{sec:time_av}

\begin{figure}
\includegraphics[scale=1.00]{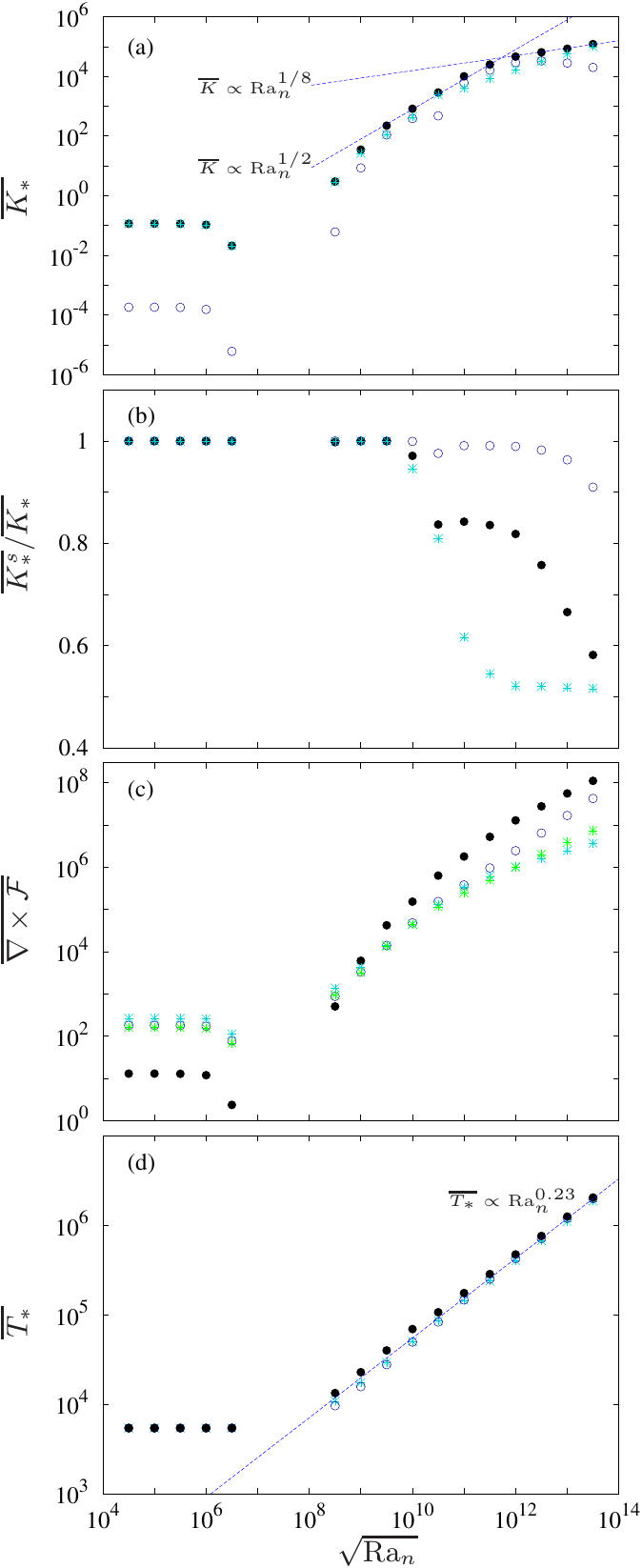}  
\caption{Physical properties versus $\sqrt{\Ray_n}$: (a) Time-averaged
  global (volume-averaged) kinetic energies $K$ ($\bullet$), $K_{a}$
  ($\circ$) and $K_{na}$ ($\ast$). The fitting lines are
  $\overline{K}=8\times 10^{-8}\Ray_n^{1/2}$ and $\overline{K}=5\times
  10^{1}\Ray_n^{1/8}$. (b) Ratios of the time-averaged equatorially
  symmetric over global kinetic energies $K^{s}/K$ ($\bullet$),
  $K_{a}^{s}/K_{a}$ ($\circ$) and $K_{na}^{s}/K_{na}$ ($\ast$). (c)
  Time-averaged rms curl of forces integrals $\mathcal{\nabla\times
    F}_I$ ($\bullet$, Inertial), $\mathcal{\nabla\times F}_V$
  ($\circ$, viscous) and $\mathcal{\nabla\times F}_C$ ($\ast$,
  Coriolis). (d) Time-averaged temperatures $\bar{T_1}$, $\bar{T_2}$
  and $\bar{T_3}$ taken at $\varphi=0$, $\theta=3\pi/8$ and
  $r=r_1=r_i+0.15d$ ($\ast$), $r=r_2=r_i+0.5d$ ($\circ$) and
  $r=r_3=r_i+0.85d$ ($\bullet$), respectively (where $d=r_o-r_i$).
  The fitting line corresponds to $\overline{T_2}=(1.95\pm
  0.15)\Ray_n^{0.223\pm 0.001}$.}
\label{fig:prop_mean_Ran} 
\end{figure}

\begin{figure}
\includegraphics[scale=1.00]{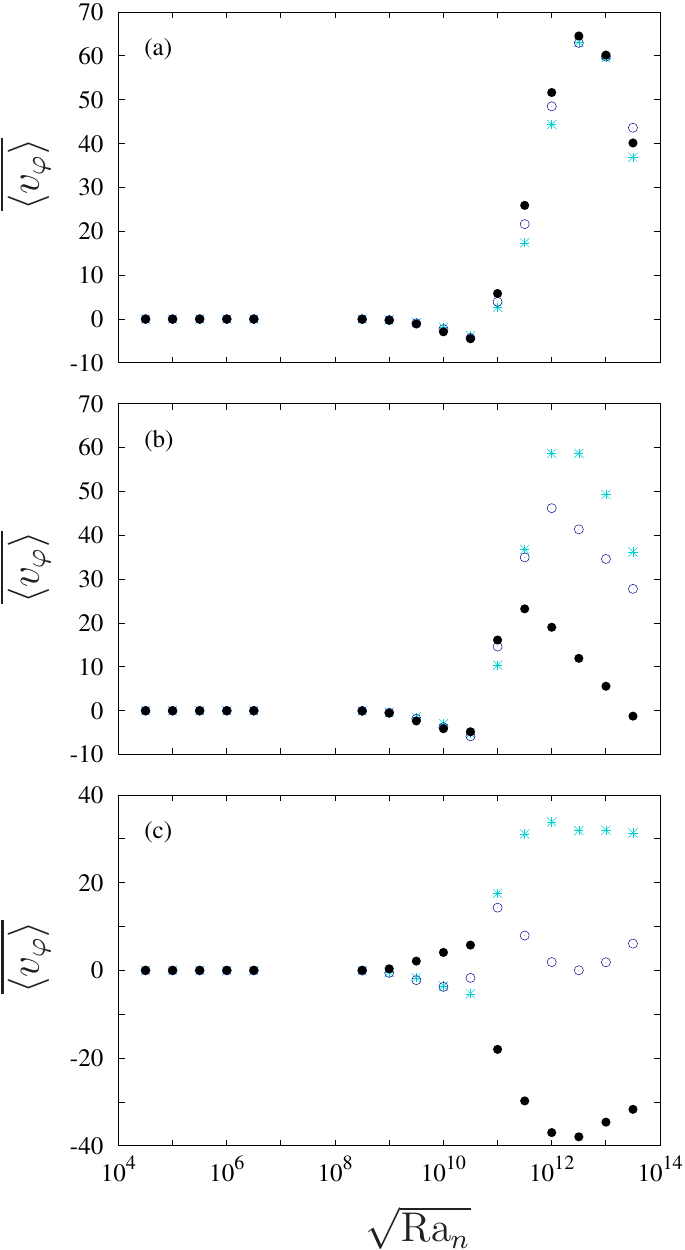}  
\caption{Time-averaged mean zonal flow (azimuthally-averaged
  $v_{\varphi}$) versus $\sqrt{\Ray_n}$: At colatitude (a)
  $\theta=\pi/8$, (b) $\theta=\pi/4$ and (c) $\theta=3\pi/8$. The
  symbols indicate the radial position: ($\ast$) $r=r_1=r_i+0.15d$,
  ($\circ$) $r=r_2=r_i+0.5d$ and ($\bullet$) $r=r_3=r_i+0.85d$ (where
  $r_i=1.5$ and $r_o=2$).}
\label{fig:zonal_mean_Ran} 
\end{figure}

The different flow regimes distinguished in
Table~\ref{table:prop_mean_ener} are investigated with the help of
time averages displayed in Fig.~\ref{fig:prop_mean_Ran} and
Fig.~\ref{fig:zonal_mean_Ran}. The former examines the volume averages
of the different components of the kinetic energy density or the
volume averaged force balance, but also temperature recorded at
different points inside the shell. The latter analyses the zonal flow
also at different colatitude and radial positions (see figure
captions). In the left region of both figures (up to $\Ray_n=10^{12}$)
the plotted variables are constant and equal to those corresponding to
the differentially heated nonburning solution at $\Ray_n=0$. As
mentioned in the previous section, at $\Ray_n=10^{13}$ the imposed
temperature gradient becomes less efficient at maintaining convection,
and the kinetic energy density of the solution decreases. At
$\Ray_n=10^{14},10^{15},10^{16}$ convection is no longer sustained, and
the triple alpha conductive state is recovered. This is why these points
are missing in the plots.

In the first burning convective regime, $10^{17}\le \Ray_n\le 10^{19}$,
corresponding to oscillatory solutions (regime I of~\cite{GWA13}), the
zonal component of the flow grows rapidly (see
Fig.~\ref{fig:prop_mean_Ran}(a)) and Coriolis forces are still
important (see Fig.~\ref{fig:prop_mean_Ran}(c)), helping to maintain
the equatorial symmetry of the flow (see
Fig.~\ref{fig:prop_mean_Ran}(b)). Zonal circulations are positive near
the outer and negative near the inner boundaries (see
Fig.~\ref{fig:zonal_mean_Ran}(c)). This also occurs at
$\Ray_n=10^{20},10^{21}$ but these solutions are different because the
equatorial symmetry has been broken, the Coriolis forces start to be
of second order (see Fig.~\ref{fig:prop_mean_Ran}(b) and
Fig.~\ref{fig:prop_mean_Ran}(c)), and the axisymmetric flow component
starts to decrease (see table~\ref{table:prop_mean_ener}). These
solutions belong to a transition regime across which the structure of
the mean zonal flow is strongly changed. A very similar regime was
also found in~\cite{GWA13} (where it was called the transitional regime), but
only for strongly stratified anelastic models. According
to~\cite{GWA13}, because of the strong stratification, different force
balances are achieved at different depths in the shell. In this regime
buoyancy dominates near the outer boundary while the Coriolis force is still
relevant in the deep interior. This reduces the amplitude of the zonal
flow and leads to a characteristic dimple in the center of the
equatorial jet~\cite{GWA13}. As will be shown in
Sec.~\ref{sec:cont_pl} the signature of this dimple is also present in
our unstratified models.

The second regime (also regime II in~\cite{GWA13}) corresponds to flows
($10^{22}\le\Ray_n\le10^{24}$) which have the maximum zonal component (see
table~\ref{table:prop_mean_ener}). These solutions are characterised
by large scale convective cells, volume-averaged kinetic energy
growing as $\Ray_n^{1/2}$, and inertial forces becoming more relevant with
respect to Coriolis and viscous forces. In contrast to the previous
regime the zonal flow pattern is reversed and becomes negative near
the outer boundary and positive near the inner. In this regime, the
positive mean zonal flow near the inner boundary is strongly increased
at mid and high but also close to equatorial latitudes. In these
latitudes the positive mean zonal flow is weaker. The radial
dependence is enhanced with increasing latitude. In addition, although
equatorial symmetry of the nonaxisymmetric flow is clearly broken, this
is not the case of the axisymmetric (zonal) flow.

A third regime (also regime III in~\cite{GWA13}) is obtained for the
largest $\Ray_n\ge 10^{25}$ explored, corresponding to strongly
convective and nonaxisymmetric turbulent flows. In this regime the
balance seems to be between inertial and viscous forces, with
volume-averaged kinetic energy slowly growing as $\Ray_n^{1/8}$, and
the volume averaged zonal component of the flow starting to slowly
lose equatorial symmetry, while remaining roughly constant. The change
of tendency of the axisymmetric component of the flow is also local,
as can be observed in Fig.~\ref{fig:zonal_mean_Ran}. Whilst in
equatorial regions the zonal flow remains roughly constant, at higher
latitudes it decreases quite sharply, which may be indicative of the
progressive loss of equatorial symmetry. In the case of the
nonaxisymmetric flow, its equatorially symmetric and antisymmetric
components are balanced, as reflected by the constant ratio $K^s_{na}/K_{na}\approx
0.5$. A fourth regime can be guessed by looking
Fig.~\ref{fig:prop_mean_Ran}(b). It will correspond to fully isotropic
turbulence characterized by nearly equal equatorially symmetric and
antisymmetric components of the zonal flow
($\overline{K^s_a}/\overline{K_a}\approx 0.5$).

\subsubsection{Force Balance}
\label{sec:fo_bal}

For all the $\Ray_n$ explored, the time averages of the temperature
are quite similar inside the shell
(Fig.~\ref{fig:prop_mean_Ran}(d)). The latter figure shows that mean
temperatures are slightly larger close to the outer boundary, except
for the turbulent solutions where the radial dependence seems to
disappear. Mean temperatures close to the equatorial region in the
middle of shell grow as $\overline{T}=(1.95\pm 0.15)\Ray_n^{0.223\pm
  10^{-3}}$ and thus the different flow regimes cannot be clearly
identified. Notice that the latter scaling is better satisfied at the
largest $\Ray_n$. For smaller forcing $\Ray_n\le 10^{23}$, the
exponent of the power law is a little bit larger, around $0.24\approx
1/4$. As argued in the following this exponent, and $\Ray_n^{1/2}$ and
$\Ray_n^{1/8}$ found for the kinetic energy density, can be deduced
from the equations. The results are compared with~\cite{GWA16}, a very
exhaustive and recent study about scaling regimes in spherical shell
rotating convection.

Assuming that in the energy equation (Eq.~\ref{eq:temp_Ra_n}) the
internal triple-alpha heat generation is balanced with the temperature
Laplacian $\nabla^2 T\sim \Ray_n T^{-3}e^{-\Bn/T}$, the
characteristic length scale remains roughly constant (of the same
order) with $\Ray_n$ (see Table~\ref{table:prop_mean_ener}), and
$T\sim \Ray_n^{\xi}$ with $\xi>0$ gives $\Ray_n^{\xi}\sim
\Ray_n^{(1-3\xi)}$, because if $\Ray_n$ is large the term
$e^{-\Bn/T}\sim e^{-\Bn/\Ray_n^{\xi}}\sim 1$. Then $\xi\approx 1/4$, in
very good agreement with our results. In addition, for $\Ray_n\le
10^{23}$ we have found from the simulations $\overline{K}\sim
\Ray_n^{1/2}$, i.e $U\sim\Ray_n^{1/4}$ with $U$ being the characteristic
velocity of the fluid. When viscous and Coriolis forces are balanced
with the Archimedes (buoyancy) force in the momentum equation (the
so-called VAC balance, see~\cite{GWA16} and references therein), as
Fig.~\ref{fig:prop_mean_Ran}(c) suggests, and with the previous
assumptions for the length scales, the characteristic velocity is
$U\sim T\sim\Ray_n^{1/4}$.

When the VAC balance is lost for the turbulent solutions
$\Ray_n\gtrsim 10^{24}$, a new power law $\overline{K}\sim
\Ray_n^{1/8}$, i.e $U\sim\Ray_n^{1/16}$, is obtained. This very low
exponent may be explained by again making use of the energy equation and
assuming that its inertial term starts to play a role, giving rise to a
new balance. The length scales for the temperature $L$ are no longer
constant, modifying slightly the $\nabla^2 T\sim \Ray_n
T^{-3}e^{-\Bn/T}$ balance to $T/L^2\sim\Ray_n T^{-3}$. Because the
solutions have very large velocities we may assume $\ve\cdot\nabla
T\sim \Ray_n T^{-3}e^{-\Bn/T}$, i.e, $UT/L\sim\Ray_n T^{-3}$ and thus
$UT/L\sim T/L^2$ or equivalently $UL\sim 1$. If $T\sim \Ray_n^{\xi}$
and $U\sim \Ray_n^{\zeta}$ then $\zeta=(1-4\xi)/2$ which is in close
agreement with our results. With $\xi=0.22$ we obtain
$\zeta=0.06\approx 1/16$.

Very similar scaling regimes have been reported previously in the
context of differentially heated convection~\cite{GWA16}. They are the
weakly non-linear regime (VAC balance) and the non-rotating regime (IA
balance). This may be an indication that very similar mechanisms 
govern both differentially or internally heated systems, even in
the case of temperature dependent internal sources.

\subsubsection{Triple-alpha Convective Patterns}
\label{sec:cont_pl}

\begin{figure*}
\includegraphics[scale=1.15]{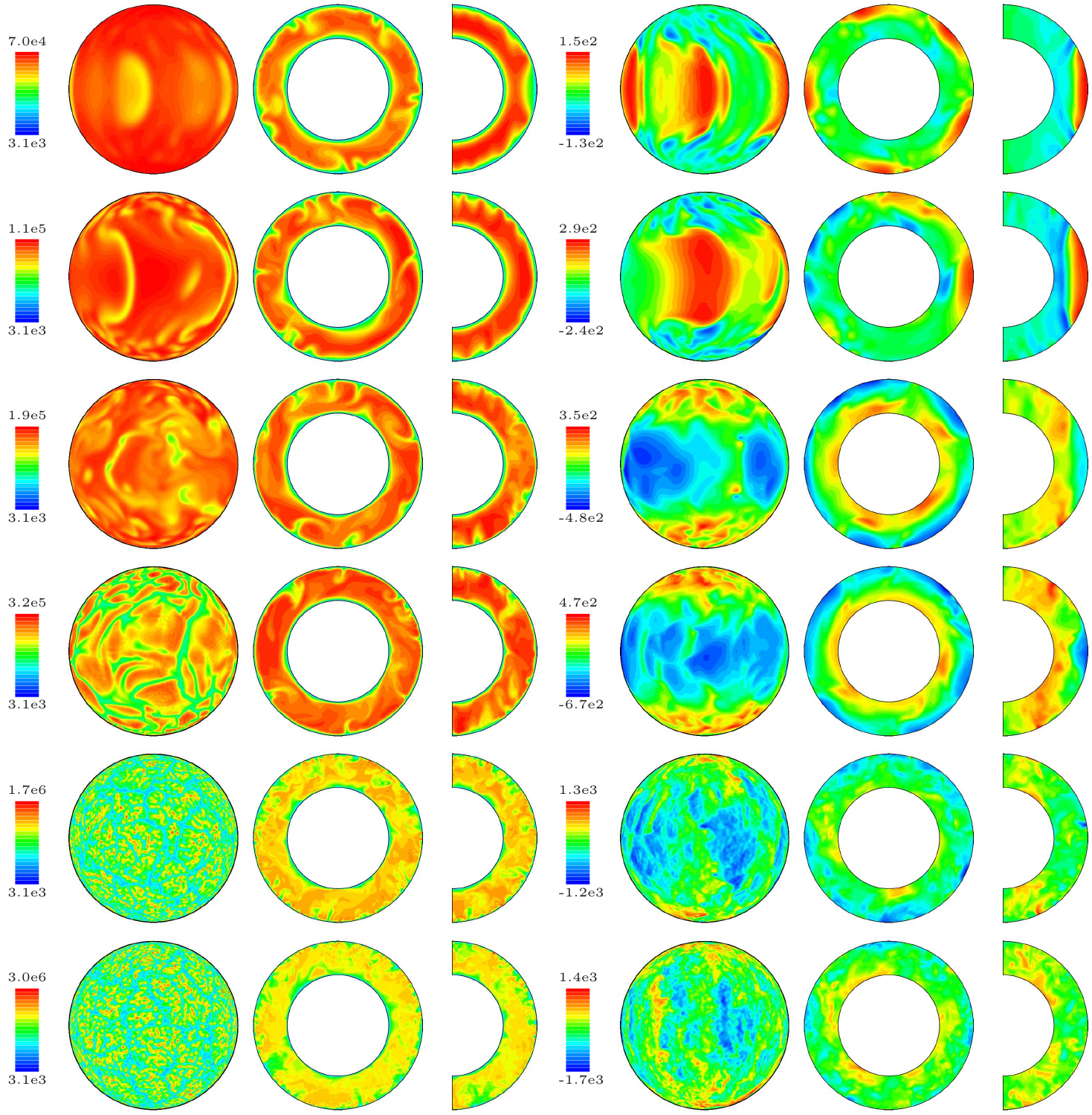}  
\caption{Instantaneous contour plots at
  $\Ray_n=10^{20},10^{21},10^{22},10^{23},10^{26},10^{27}$ (from top
  row to bottom row). The left three plots are the spherical,
  equatorial and meridional cross-sections of the temperature $T$. Right
  three plots: Same cross-sections for $v_{\phi}$. Meridional cross-sections and
  the spherical cross-section of $T$ are taken at a relative maximum. The
  spherical cross-section of $v_{\phi}$ is at the outer surface.}
\label{fig:cplt_Ran}
\end{figure*}

\begin{figure*}
\includegraphics[scale=1.15]{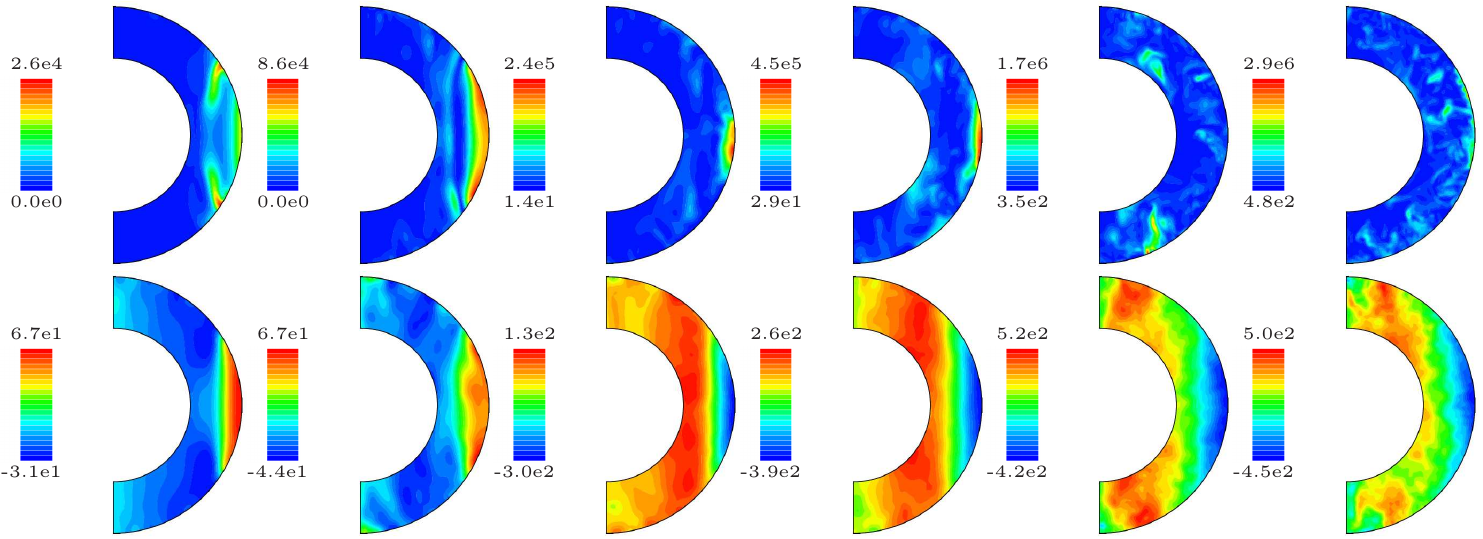}  
\caption{Instantaneous contour plots at
  $\Ray_n=10^{20},10^{21},10^{22},10^{23},10^{26},10^{27}$ (from left
  to right) showing the meridional cross-sections of the kinetic energy
  density $\ve^2/2$ (1st row) and the mean zonal flow
  (azimuthally-averaged $v_{\varphi}$) (2nd row). Cross-sections of
  $\ve^2/2$ pass through a maximum.}
\label{fig:cplten_Ran} 
\end{figure*}

The variation of the topology of the flow with increasing
$\Ray_n=10^{20},10^{21},10^{22},10^{23},10^{26},10^{27}$, i.e the
patterns of triple alpha convection, can be analysed with the help of
Fig.~\ref{fig:cplt_Ran} and Fig.~\ref{fig:cplten_Ran}. The former
displays the contour plots of the temperature (with increasing
$\Ray_n$ from top to bottom rows) in spherical, equatorial and
meridional cross-sections (left group of plots), as well as those for the
azimuthal velocity (right group of plots). The latter compares the
total kinetic energy of the flow (first row) with the axisymmetric
component of the azimuthal velocity (second row) on meridional
cross-sections for the same sequence of $\Ray_n$ as
Fig.~\ref{fig:cplt_Ran}. As usual, spherical and meridional cross-sections
pass through a relative maximum. The azimuthally averaged azimuthal
velocity contour plots shown in the second row of
Fig.~\ref{fig:cplten_Ran} can be easily compared with Figure 2
of~\cite{GWA13}, showing a good agreement.

\begin{figure*}
\includegraphics[scale=1.15]{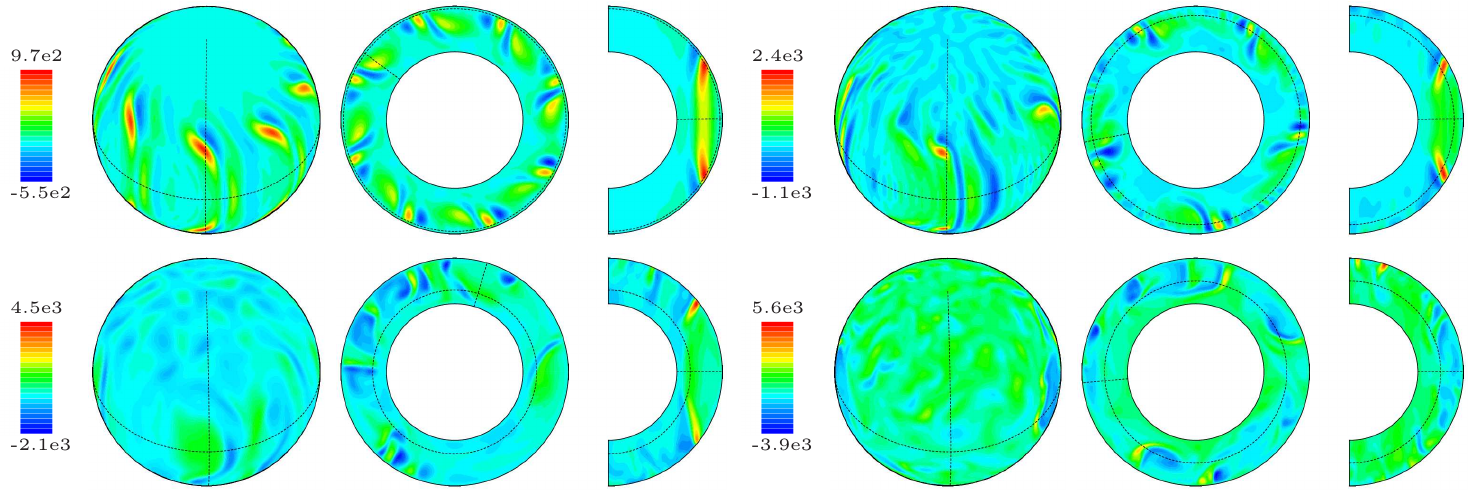}  
\caption{Instantaneous contour plots at $\Ray=6.2\times 10^3$ and
  $\Ray_n=10^{19},10^{20},10^{21},10^{22}$ (from left to right and
  from top to bottom group of plots). Each group of three plots are
  the spherical, equatorial and meridional cross-sections of the axial
  vorticity $w_z$. Spherical cross-sections are taken at $r=r_i+0.95d$,
  $r=r_i+0.8d$, $r=r_i+0.3d$ and $r=r_i+0.5d$ for
  $\Ray_n=10^{19},10^{20},10^{21},10^{22}$, respectively. Meridional
  cross-sections are taken where $w_z$ has a relative maximum.}
\label{fig:cpltvor_Ran}
\end{figure*}

At $\Ray_n=10^{20}$ the solutions belong to the oscillatory
regime. Temperature is maximum in the polar regions and almost
conductive (see spherical cross-section). A remarkable characteristic of the
burning convection is that large azimuthally elongated outward plumes
coexist with very narrow inward cold cells. The outward plumes are
mushroom like and correspond to a large outward radial flow. At lower
latitudes, close to the equatorial region, an $m=5$ convective pattern
develops (see equatorial cross-section). This $m=5$ pattern is clearly
distinguished in the contour plots of $v_{\varphi}$. There is an
equatorial belt of positive zonal circulation attached to the outer
boundary. At higher latitudes the zonal circulation is strongly
negative, decreasing its amplitude up to the poles. The total kinetic
energy density, see meridional cross-section shown in
Fig.~\ref{fig:cplten_Ran} (first row, left), is maximum just where the
negative circulation takes place, while the axisymmetric azimuthal
velocity (Fig.~\ref{fig:cplten_Ran} second row, left) is maximum at
the positive equatorial belt. This figure reveals also the strong
equatorial symmetry of the flow.

Very narrow downwelling plumes, elongated in the colatitude direction,
are clearly observed on the equatorial region at $\Ray_n=10^{21}$ as
Coriolis forces are still noticeable (see temperature spherical
cross-sections of Fig.~\ref{fig:cplt_Ran}). Colatitude directed
coherent vortices have been also obtained in Boussinesq (Fig. 4
of~\cite{HeAu12}) or anelastic (Fig. 6 of~\cite{BBBMT08}) models, in
the contexts of Saturn's atmosphere and solar convection,
respectively. The latter models considered differential heating in a
regime strongly influenced by rotation. We note however the very
different azimuthal and radial nature of our coherent vortices when
considering an additional source of internal heating. The flow
patterns at $\Ray_n=10^{21}$ are similar as those at $\Ray_n=10^{20}$
but with a weaker zonal component (see 2nd plot, 2nd row of
Fig.~\ref{fig:cplten_Ran}). The positive equatorial belt has been
disrupted, becoming a large convective cell in which the kinetic
energy is concentrated (see 2nd plot, 1st row of
Fig.~\ref{fig:cplten_Ran}). The flow then has strong azimuthal
symmetry $m=1$ (see the equatorial cross-sections of
Fig.~\ref{fig:cplten_Ran} and
table~\ref{table:prop_mean_ener}). Convection is spreading to high
latitudes, and polar zonal circulations become important (see 2nd
plot, 2nd row of Fig.~\ref{fig:cplten_Ran}). The flow is still
equatorially symmetric ($K_s>0.8K$) but antisymmetric motions increase
noticeably when compared with solutions at lower $\Ray_n$. The dimple
seen in the equatorial belt of zonal motions (see 2nd plot, 2nd row of
Fig.~\ref{fig:cplten_Ran}) was associated in~\cite{GWA13} with a
transitional regime. As we argue at the end of this section, this
regime seems also be valid in our case.

The corresponding flow topology at $\Ray_n=10^{22},10^{23}$ is quite
different from that at lower $\Ray_n$. While large scale
temperature vortices are still present, convection is more vigorous and more
developed, with small scale motions in the polar regions. The
meridional and latitudinal extent of the convective cells is starting
to decrease (see spherical cross-sections of $T$) while it remains large
(similar to the gap width) in the radial direction (see equatorial
cross-sections of $T$). This gives more support to our assumption of a
characteristic length scale of order unity when deriving the
corresponding temperature and velocity scalings in
Sec.~\ref{sec:fo_bal}. As stated, when studying time averaged
properties from $\Ray_n\ge 10^{22}$, the flow pattern has reversed and
now the circulation occurring in the equatorial belt is negative
whilst at higher latitudes it is positive. The meridional extension of
the equatorial belt decreases, and the motions are more attached to the
outer surface (see 3nd/4th plot, 1st and 2nd row of
Fig.~\ref{fig:cplten_Ran}). While the axisymmetric flow (and thus the
equatorial belt) is still equatorially symmetric, this symmetry is
progressively lost on the rest of the shell. When comparing to the
corresponding differentially heated Boussinesq models of Fig. 2
in~\cite{GWA13}, our models exhibit larger steady axially oriented
regions (green color), which tend to spread to larger latitudes, i.e
moving radially inwards, with increasing $\Ray_n$.

The two most extreme cases at $\Ray_n=10^{26},10^{27}$ exhibit very
fine temperature structures close to the outer boundary, and of
granular type like those observed on the Sun's surface. Nevertheless, the
large scale weak temperature vortices still recall the topology of the
triple alpha conductive state. The negative and positive circulations
become more nonaxisymmetric, the latter nearly reaching the
poles. Although the number of small scale structures is increasing,
negative azimuthal velocity cells, elongated on the meridional
direction, are still present on equatorial latitudes even at
$\Ray_n=10^{27}$. In the latter case, these negative meridional cells
alternate with very thin positive plumes that connect positive
circulations of both poles (see spherical cross-section of
$v_{\varphi}$). Although motions are concentrated very close to the
outer boundary (see rightmost plots, 1st row of
Fig.~\ref{fig:cplten_Ran}), they are no longer located near equatorial
latitudes and may reach high latitudes. In addition the equatorial
symmetry of the flow is clearly lost, although not that of its
axisymmetric component (see also rightmost plots, 2nd row of
Fig.~\ref{fig:cplten_Ran}). The latter is stronger at high latitudes,
in contrast to what happens at lower $\Ray_n$.

The typical anelastic transitional flows of~\cite{GWA13} exhibit
quasigeostrophic structures in the interior of the shell, whilst they
are more buoyancy dominated (radially aligned) near the outer
boundary. The critical radial distance, $r_{\text{mix}}$
in~\cite{GWA13}, separating the two dynamical behaviours within the
shell increases (from the outer boundary to the inner) with the
forcing. This also occurs in our Boussinesq models, as shown in
Fig.~\ref{fig:cpltvor_Ran}, but with different patterns which are due to
the different forcing mechanisms (differentially heated anelastic
versus triple-alpha internal heating) of the convection. The figure
displays the contour plots of the axial vorticity $w_z$ on
spherical, meridional and equatorial cross-sections (the latter are shown in
Fig. 13 of ~\cite{GWA13} for three anelastic models). The positions of
the cross-sections are displayed with black lines in the contour plots. Four representative
solutions belonging to regimes I, transitional, and II, at
$\Ray_n=10^{19},10^{20},10^{21},10^{22}$, are shown.  In regime I
the flow is quasigeostrophic outside the tangent cylinder while
convection is absent in the inner region (see left top group of
plots). At $\Ray_n=10^{20}$, a rough critical radius
$r_{\text{mix}}=r_i+0.8d$ can be identified (see right top group of
plots, equatorial and meridional cross-sections), becoming larger
$r_{\text{mix}}=r_i+0.3d$ at $\Ray_n=10^{21}$ (see left bottom group
of plots). In agreement with~\cite{GWA13}, for $r>r_{\text{mix}}$
vorticity is radially aligned (best seen in the meridional cross-sections) while
for $r<r_{\text{mix}}$ the quasigeostrophic columns can be
identified. However, in contrast to~\cite{GWA13} this behaviour is
more evident in the polar regions (see spherical cross-sections) than at
lower latitudes where axially aligned structures at
$r>r_{\text{mix}}$ still survive. For flows belonging to regime II small
convective structures within all $r$ can be found (see right bottom
group of plots) but they are now quite aligned in the vertical
direction because the negative zonal circulations at low latitudes are
becoming stronger.

\begin{table*}[t]
$$
\begin{array}{lcccccccccccc}
\hline  
\Ray_n  & f^{T}_{\text{max}} \  & [f^{T}_{1},f^{T}_{2}] & \overline{f^T} & f^{v_{\varphi}}_{\text{max}} & [f^{v_{\varphi}}_{1},f^{v_{\varphi}}_{2}]& \overline{f^{v_{\varphi}}} & f^{\langle v_{\varphi}\rangle }_{\text{max}} & [f^{\langle v_{\varphi}\rangle }_{1},f^{\langle v_{\varphi}\rangle }_{2}] & \overline{f^{\langle v_{\varphi}\rangle}}\\  
\hline
10^{17} & 8.77             & [8.77,80.0]    &  9.21     & 8.77                 & [8.77,80.0]                           & 10.6                      & --                                         &  --          & --  \\
10^{18} & 11.8             & [11.8,59.1]    &  14.8     & 11.8                 & [11.8,71.2]                           & 15.0                      & --                                         &  --          & --  \\
10^{19} & 21.0             & [2.50,31.0]    &  15.9     & 21.0                 & [5.29,41.9]                           & 22.1                      & 34.57                                      & [1.23,45.0]  & 25.2\\
10^{20} & 28.2             & [2.17,39.6]    &  24.4     & 28.2                 & [2.54,56.3]                           & 26.5                      & 8.881                                      & [0.73,52.1]  & 32.8\\
10^{21} & 16.1             & [3.81,60.3]    &  61.1     & 12.7                 & [7.00,66.4]                           & 30.1                      & 6.44                                       & [1.99,55.9]  & 45.8\\
10^{22} & 26.0             & [14.7,91.1]    &  171      & 12.5                 & [2.03,57.2]                           & 71.2                      & 2.38                                       & [7.52,80.4]  & 49.7\\
10^{23} & 16.7             & [16.7,235]     &  345      & 4.58                 & [4.58,56.5]                           & 38.4                      & 2.48                                       & [2.48,90.2]  & 111\\
10^{24} & 82.8             & [25.1,287]     &  375      & 45.6                 & [4.22,102]                            & 90.5                      & 31.53                                      & [18.0,66.2]  & 35.0\\
10^{25} & 54.3             & [23.8,360]     &  270      & 17.4                 & [13.9,128]                            & 180                       & 33.80                                      & [5.62,115]   & 50.5\\
10^{26} & 110              & [16.8,372]     &  402      & 91.8                 & [6.70,252]                            & 377                       & 19.91                                      & [19.9,164]   & 179\\
10^{27} & 42.3             & [16.9,509]     &  948      & 35.3                 & [14.1,350]                            & 233                       & 47.48                                      & [10.6,231]   & 198\\
\hline
\end{array}					  
$$
\vspace{-0.3cm}
\caption{Burning Rayleigh number $\Ray_n$, the frequency with larger
  amplitude $f^{*}_{\text{max}}$, the interval $[f^*_1,f^*_2]$
  containing the 10 first frequencies with larger amplitude, and mean
  frequency $\overline{f^{*}}$ of the frequency spectrum of
  temperature $T$, azimuthal velocity $v_{\varphi}$ and zonal flow
  $\langle v_{\varphi} \rangle$ taken close to the outer boundary and
  the equatorial plane (at
  $(r,\theta,\varphi)=(r_i+0.85d,3\pi/8,0)$).}
\label{table:freq}
\end{table*}

\subsubsection{Flow Time Scales}
\label{sec:flw_ts}

It is interesting to address which are the relevant time scales
present in the different flow regimes obtained with the
DNS. Table~\ref{table:freq} lists the frequency with larger amplitude
$f^{*}_{\text{max}}$, the interval $[f^*_1,f^*_2]$ containing the 10
first frequencies with larger amplitude, and the mean frequency
$\overline{f^{*}}$ of the frequency spectrum of temperature $T$,
azimuthal velocity $v_{\varphi}$ and zonal flow $\langle v_{\varphi}
\rangle$ taken close to the outer boundary and the equatorial plane
(at $(r,\theta,\varphi)=(r_i+0.85d,3\pi/8,0)$). Rotating waves
($\Ray_n=10^{17},10^{18}$) close to the onset have time scales quite
similar to the critical frequencies $2\pi/\omega_c\sim 10$. Usually
secondary bifurcations giving rise to oscillatory flows involve large
time scales, as is the case for $\Ray_n=10^{19},10^{20}$, with relevant
frequencies $f\sim 1$.

With increasing $\Ray_n\ge 10^{21}$, time scales that are several orders of magnitude smaller become important, but nonetheless larger time scale
similar to those at the onset remain relevant: while $f^*_1\sim 10$
remains quite constant $f^*_2\sim 10^2$ increases, but not too
much. This also occurs in differentially heated convection with
increasing $\Ray$. Thus the study of the onset of convection is of
fundamental importance, because it reveals time scales that are found to
be present even at highly supercritical regimes. Going into more detail on the latter, by comparing $\overline{f^T}$ with
$\overline{f^{v_{\varphi}}}$ we see that the flow has larger time
scales than the temperature variations. This is particularly true in the case
of the zonal flow having the smallest $\overline{f^{\langle
    v_{\varphi}\rangle}}$ which are very similar to the conditions
corresponding to the onset. For instance, at $10^{19}\le \Ray_n \le
10^{25}$ time scales are $\overline{f^{\langle
    v_{\varphi}\rangle}}\lesssim 50,$ while at the onset they are
$2\pi/\omega_c\approx 10$. We recall that zonal flow time scales are
important because of their observational consequences.

\section{Convection in Accreting Neutron Star Oceans}
\label{sec:astr}

Before linking our results to convection occurring in accreting
neutron star oceans (see the introductory section) we should stress
that we are still far from realistic modelling. In such environments
the flow is stratified, thus compressible convection should be
considered. However, as argued in~\cite{GCW18}, Boussinesq convection
constitutes the very first step towards the study of convective
patterns in global domains. Some insights into the consideration of
compressibility effects are provided in~\cite{GWA13}, by means of the
anelastic approximation. According to this study the same I,II and III
flow regimes, including the transitional regime (between I and II),
are obtained with strong stratification. Retrograde zonal flow
amplitudes are decreased with respect to the Boussinesq case in regime II
 but this seems not to happen in regime I. In addition,
strongly stratified flows favour the appearance of the transitional
regime, and the transition to regime III is delayed.

As mentioned before, this study is focussed primarily on the effect of
increasing internal heating via triple-alpha reactions, so the values
chosen for $\Pr$, $\Ta$, and $\eta$ are not extreme.   In addition, in real
stars the fraction of helium would deplete over time as carbon and
heavier elements form - while in this study, for fixed $\Ray_n > 0$,
heat is generated continuously in the ocean. More realistic models,
accounting for nuclear physics as well as hydrodynamics, would require
coupling the Navier-Stokes equation, energy equation, and a nuclear
reaction network; far outside the scope of this initial study.

When helium is burning steadily, (see~\cite{Bil98} for details) heat
generated by nuclear reactions is balanced by radiative cooling
($q_{\text{cool}} \propto T^4$). When this balance no longer, holds a
thermal instability gives rise to a thermonuclear runaway and the
associated Type-I X ray burst~\cite{Bil98}.  In the latter paper,
marginal stability curves for several accretion rates were given as a
function of temperature and column depth; these calculations indicated
that unstable burning (bursting) should continue up to accretion rates
a factor of a few larger than the value inferred from observations.
Recent studies (see~\cite{ZCN14} and the references in the
introduction) have considered an additional source of heat at the base
of the ocean, to bring the critical accretion rate at which burning
stabilises closer to observational values. Other studies~\cite{KLZ09},
based on one-zone hydrodynamic stellar evolution models, have achieved
a similar effect by considering an inward diffusion of helium due to
the influence of rotation and magnetic fields. In contrast to these
studies, in our study nuclear heating is balanced by both thermal
advection and diffusion.  For the latter we are assuming constant
density and thermal conductivity (Boussinesq), and any additional sink
of heat is located at the upper boundary. We are thus not taking into
account radiative processes\footnote[1]{In stellar interiors the heat
  flux is also radiative (see for instance~\cite{BrBi98}) giving rise
  a total thermal conductivity
  $k_{\text{tot}}=4acT^3/3\kappa_{\text{rad}}\rho+k_{\text{cond}}$,
  with $a$ the radiation density constant, $c$ the speed of light,
  $\kappa_{\text{rad}}$ the radiative opacity and $k_{\text{cond}}$
  the thermal conductivity of Fourier's law of heat conduction. We
  have only considered $k_{\text{tot}}=k_{\text{cond}}$ in our
  modelling.} but focusing on the heat convective transport on a
rotating spherical shell.

There are several numerical studies (local in nature) modelling 
convection during bursts on accreting neutron stars (see for
instance~\cite{WHCHPRFSBW04,Malone11}). In the former, convection is
implemented using mixing length theory, while in the latter the
momentum equation (advection-like) for convective velocity
is solved.  These studies (~\cite{WHCHPRFSBW04,Malone11})  - but also
earlier ones~\cite{WZW78} devoted to the study of stellar evolution
of massive stars - rely on the Schwarzschild and Ledoux criterion to compute
the onset of convection (when thermal and compositional buoyancy
overcomes gravity). In the Schwarzschild criterion an instantaneous
chemical equilibrium is assumed, which favours the appearance of the
convective instability~\cite{WHTR04}. This means that the fluid can be
Schwarzschild unstable but Ledoux stable to velocity perturbations, referred to as semiconvection in previous
studies~\cite{WZW78,WHCHPRFSBW04,Malone11}. These studies (and
also~\cite{WBS06,YuWe18}) have built up the current picture of the
role played by convection in the thermal evolution of a burst. Burning
is mainly localised at the base of the accreted layer where
temperature and density are larger. Because the burning time scale is much
shorter than the time scale of radiative processes, nuclear energy is more
efficiently dissipated by convection. Fluid motions spread towards the
upper radiative layers of the accreted ocean until the entropy of the
convective layer balances the radiative entropy.  Finally, because of
the decrease of the burning rate, for instance due to fuel consumption
in the nuclear reaction chain (see Fig. 9 of~\cite{KeHe17} or Fig. 22
and 24 of~\cite{WHCHPRFSBW04}), or the mixing of fuel with other
elements, convection dies away - at which point the radiative flux
starts to transfer the energy out to the photosphere. In some bursts
the observed luminosity exceeds the Eddington limit, giving rise to the
phenomenon of photospheric radial expansion (PRE) in which a radiation
driven wind pushes the photospere outwards, ejecting both freshly-accreted matter
and heavy-element ashes~\cite{YuWe18}.

Using this picture as a guide, we can try to model the convective
evolution of a burst by varying $\Ray_n$. The physical meaning of
varying this parameter was described in Sec.~\ref{sec:model}.  Varying
$\Ray_n$ could be associated with a variation in the size $d$ of the
convective layer or the variation of helium mass fraction. For
simplicity, we assume the latter and try to mimic the evolutionary
history of the pure helium flash model of~\cite{WHCHPRFSBW04}.  We
note that $\Ray$ would also change during a burst; however because
$\Ray$ is much less temperature-sensitive, its variations are
neglected in our model. In the early stages of the burst we assume
that $\Ray_n$ is subcritical and thus nuclear heat is simply
transported by conduction ($q_n\sim \nabla^2 T$). Because of
accretion, $\Ray_n$ increases, reaching supercritical values and
convective motions start to contribute to dissipate the heat of the
layer (the advection term $\ve\cdot\nabla T$ in the energy equation
(Eq.~\ref{eq:temp_Ra_n}) increases). Notice that the advection term
plays the same mathematical role in the energy equation than the
additional heating term introduced in the motionless models
of~\cite{ZCN14}. A heat flux from the bottom surface was assumed, and
provided better predictions of critical accretion rates than those
using the usual approach of balancing only nuclear heating with
radiative cooling~\cite{Bil98}.

\begin{table}[b!]
$$  
\begin{array}{lccccccccccccccc}
\hline  
d~(\text{cm})& r_o~(\text{cm})  & \Omega~(\text{s}^{-1}) & g~(\text{cm} ~\text{s}^{-2})\\[2.pt]
10^{2} & 10^{6} &  10^{3} & 10^{14} \\
\hline
\gamma~(\text{s}^{-2}) &\nu~(\text{cm}^{2} ~\text{s}^{-1}) & \kappa~(\text{cm}^{2} ~\text{s}^{-1}) & \alpha~(\text{K}^{-1})\\[2.pt]
 10^8 & 10^0 & 10^{4} & 10^{-7}\\
\hline  
C_p~(\text{cm}^2 ~\text{s}^{-2} ~\text{K}^{-1}) & \Delta T~(\text{K}) & T_o~(\text{K}) & \rho~(\text{g} ~\text{cm}^{-3})\\[2.pt]
 10^9 & 2\times 10^{8}  & 10^{8} & 10^7\\
\hline
\end{array}
$$
\caption{Estimated physical properties of a pure Helium ocean
  from~\cite{Yakovlev80,NaPe84,Wat12} (see also~\cite{GCW18})}
\label{table:phys_prop}
\end{table}

Because at some point in the rising phase of the burst convection
starts to recede back inwards~\cite{WHCHPRFSBW04} some criterion to
start decreasing $\Ray_n$ down to subcritical regimes must be
assumed. Our criterion is that the advection term becomes as large as
nuclear burning rate and cooling term in the energy equation. That is
$\ve\cdot\nabla T\sim q_n\sim \nabla^2 T$ which is in some sense
similar to the entropy balance assumed in~\cite{WBS06} (when computing
the radial extent of the convective layer). The model of~\cite{WBS06}
neglects compositional changes due to convection and the advection term
in the entropy equation when deriving the thermal state of the
convective layer. In our modeling, at the flow regimes I and II (see
Sec.~\ref{sec:fo_bal}) the advection term can be neglected, and
our entropy equation becomes similar to theirs.

The limit $\ve\cdot\nabla T\sim q_n\sim \nabla^2 T$ is defined by some
critical $\Ray_n$ at the boundary between flow regimes II and III
(see Sec.~\ref{sec:fo_bal}).  These flows have 
interesting properties similar to those the global buoyant
$r$-modes which arise from perturbations of the tidal
equations~\cite{Hey04}, and which have been suggested as good candidates to explain the observed
burst oscillations in the cooling phase (tail) of the
burst. According to~\cite{Hey04}, the buoyant $r$-modes have low wave number,
are retrograde (propagating westward) and span a wide colatitudinal
region around the equator. The flow characteristics shown
Sec.~\ref{sec:cont_pl} for the II regime are quite similar: strong
outer retrograde circulations near the equator with relatively strong
low wave number azimuthal symmetry. By departing from this type of
flow, and decreasing $\Ray_n$, our model predicts a decrease in
corotating frame frequency (see Sec.~\ref{sec:flw_ts}).  This 
 agrees qualitatively with the observed drift of the frequency of burst
oscillations. The convergence of burst oscillation frequency towards a value close to the neutron
spin frequency~\cite{Wat12} seen in the tail of many bursts might be
explained by the fact that as very low $\Ray_n$ is reached,
convection is no longer influenced by burning but is instead differentially heated, so that the limit frequency corresponds to that estimated for
differentially heated systems~\cite{GCW18}.

\begin{table*}[t!]
\begin{center}
\begin{tabular}{lcccccccccccc}
\hline\\[-8pt]
$\rho~(\text{g}~\text{cm}^{-3})$& & $y~(\text{g}~\text{cm}^{-2})$& & $\Ray$& & $\Ray_n$& & $\Bn$& & $T_2 ~(\text{K})$& & $U^*~(\text{m}~\text{s}^{-1})$ \\
\hline\\[-8pt]
$10^7$& & $10^9$& & $4\times 10^{13}$& & $9\times 10^{77}$& & $9\times 10^{18}$& & $2\times 10^{8}$& & $7\times 10^{1}$ \\
\hline
\end{tabular}
\end{center}                                                                                                                         
\caption{Some estimations for a helium accreting neutron star ocean.}
\label{tab:ns_param}
\end{table*}

In the study by \cite{Spitkovsky02} of the regimes for the spreading of a nuclear burning
front, lateral shear convection was driven by
inhomogeneous radial expansion, and rotation was taken into
account. The authors of~\cite{Spitkovsky02} conjectured that
inhomogeneous cooling (from the equator to the poles) might drive strong
zonal currents that could be responsible for burst oscillations. The regimes
described in~\cite{Spitkovsky02} agree qualitatively quite well with
our results for flow regime II. Convection takes the
form of strong lateral shear retrograde zonal flows, and inhomogeneous
cooling is present.

Here we are trying to simulate the progress of convection during a Type I X-ray burst by changing the value 
$\Ray_n$. Helium burning due to accretion is modelled with an increase
of $\Ray_n$, whilst this parameter is strongly lowered when a
thermonuclear runaway develops. This means that our convective
models saturate (reach the statistically steady state) faster than the
rate of change we are assuming for $\Ray_n$. In other words, flow
saturation time scales should be significantly shorter than the time
scales associated with accretion processes and the time to reach
 peak luminosity (seen in the X-ray lightcurves) after helium has
ignited. Because the arguments above are quite speculative, and our
model does not incorporate much of the relevant physics that occurs in accreting
neutron star oceans, we are quite far from a realistic application of
the results. However, some qualitative behaviour is reproduced, motivating more in-depth research.

Some predictions for the dimensionless modelling parameters and flow
properties of an accreting neutron star ocean are provided in the
following. Estimates for the physical properties of a pure Helium ocean (see
Table~\ref{table:phys_prop}) give rise to $\Pr=10^{-4}$, $\Ta=10^{14}$
and $1-\eta=10^{-4}$ (see also~\cite{GCW18}). The estimated Rayleigh
numbers $\Ray$ and $\Ray_n$ and the estimated exponent $\Bn$ in the
triple alpha heat source are listed in Table~\ref{tab:ns_param} for
$\rho=10^7 ~\text{g} ~\text{cm}^{-3}$. The density corresponds to
column depth $y=p/g=10^9 ~\text{g} ~\text{cm}^{-2}$ shown in Fig. 1
of~\cite{Bil98} which displays the helium ignition conditions
(temperature and column deep) for several accretion rates on a typical
neutron star.  The estimated $\Ray\sim 10^{13}$ is
strongly supercritical. At $\Pr=10^{-4}$, $\Ta=10^{12}$ and
$1-\eta=10^{-1}$, the parameters closest to the values estimated for neutron stars
currently reached in the linear stability analysis of~\cite{GCW18},
the critical $\Ray$ for the onset of convection is of order
$10^3$. Then, the influence of temperature gradients (imposed between
the boundaries) on the flow dynamics becomes quite strong and
very large $\Ray_n$ (i.e triple-alpha heat sources) will be needed to
have convection driven effectively by burning and to reach
flow regime III, from which the thermonuclear runaway transfers the energy to
the photosphere. According to Table~\ref{tab:ns_param} larger $\Ray_n$
seems to be likely, with $\Ray_n>10^{77}$ (in part because of the large
pre-factor $10^{45}$ in its definition), but it is not clear if the
transition between regimes II and III will persist, and further
research is required. Our preliminary numerical explorations (not
shown in this study) have revealed that the onset of burning
convection ($\Ray_n^c$) depends on the imposed temperature gradients,
i.e, on $\Ray$, particularly if they are supercritical. Hence the
critical $\Ray_n^{\text{tran}}$ for the transition between 
flow regimes II and III may also depend on $\Ray$. The numerical estimation of
this dependence $\Ray^{\text{tran}}_n(\Ray)$ and that of
$\Ray^c_n(\Ray)$ is then relevant for the knowledge of flow regimes
(such as those described in our study) that may occur in accreting
neutron star oceans.

With the estimations for the parameters given in
Table~\ref{tab:ns_param}, the characteristic temperature $T^*$ and
velocity $U^*$ can be estimated assuming the power laws
$\overline{T_2}=1.95\Ray_n^{0.223}$ and $\overline{K}=50\Ray_n^{1/8}$
derived in the previous section for the time averages of the
temperature near the equator in the middle of the shell and the
volume-averaged kinetic energy density: specifically
$T^*=\nu^2\gamma^{-1}\alpha^{-1}d^{-4}T_2 ~\text{K}$ and $U^*=\nu
d^{-1}\sqrt{2K} ~\text{cm} ~\text{s}^{-1}$. Their estimated values are
reasonable when compared with the neutron star scenario in Fig. 1
of~\cite{Bil98}. Our estimated value $T^*=2\times 10^8~\text{K}$ at
$y=10^9 ~\text{g} ~\text{cm}^{-2}$ lies in the region in which the
ocean is thermally unstable. Although flow velocities are large,
$U^*=70~\text{m} ~\text{s}^{-1}$, this is still well below the sound
speed for a neutron star ocean~\cite{Eps88}. Our simulations suggest
that zonal motions in the case of developed burning convection would
be smaller than this.


\section{Summary}
\label{sec:conc}

This paper has carefully investigated several flow regimes of
Boussinesq convection in rotating spherical shells, driven by a
temperature dependent internal heating source. This constitutes a
first step in the understanding of convection driven by triple-alpha
nuclear reactions occuring in rotating stellar oceans. Stress-free
boundary conditions are imposed, and the parameters ($\eta=0.6$,
$Ta=2\times 10^5$ and $\Pr=1$) have been chosen since they are
numerically reasonable. They are similar, allowing comparisons, to
those of several previous studies~\cite{GWA13} of convection driven by
an imposed temperature gradient used to model planetary
atmospheres. The three-dimensional simulations presented here, which
have neither symmetry constraints nor numerical hyperdiffusivities,
provide the first numerical evidence for a notable similarity between
planetary atmospheric flows and those believed to occur in rapidly
rotating stellar oceans, which are driven by different heating
mechanisms.

For small rates of internal (nuclear-burning) heat sources (modelled
by a dimensionless parameter $\Ray_n$) a spherically symmetric
conductive state is stable. Its radial dependence differs strongly
from the differentially heated basic state. By increasing $\Ray_n$,
convection can be driven even with negative imposed temperature
gradients. In contrast to the well-known thermal Rossby waves
(spiralling columnar) preferred at moderate Prandtl
numbers~\cite{Zha92,GCW18} with differential heating, the onset of
burning convection takes place in the form of waves attached to the
outer boundary in the equatorial region. These waves are equatorially
symmetric, with very weak $z$-dependence. The frequencies do not change
substantially, nor do azimuthal wave numbers, when compared with the
differentially heated case.

By increasing $\Ray_n$ beyond a very large critical value $\Ray_n\sim
10^{17}$, a sequence of transitions leading to turbulence are
observed. This sequence takes place in quite similar fashion to the way that
differentially heated convection changes when the usual Rayleigh number is
increased~\cite{Chr02}. First travelling wave solutions, periodic in
time, are obtained. Subsequent bifurcations lead to oscillatory
convection in which the axisymmetric (zonal) component of the flow is
enhanced, and large scale motions (low mean azimuthal wave number) are
favoured. By increasing $\Ray_n$ further, the equatorial symmetry of
the solutions is broken, leading to nearly poloidal flows where the
zonal motions become less relevant.

The three different regimes of~\cite{GWA13} have been identified. In
the first regime, corresponding to oscillatory solutions, the zonal
component of the flow (positive near the outer and negative near the
inner boundaries) grows rapidly and Coriolis forces are still
important in helping to maintain the equatorial symmetry of the
flow. A second regime corresponds to solutions which have the maximum
zonal component. These solutions are characterised by large scale
convective cells, and the volume-averaged kinetic energy grows as
$K\sim\Ray_n^{1/2}$. In this regime inertial forces start to become
relevant with respect to Coriolis and viscous forces. The spatial
structure of the zonal flow is reversed, becoming negative near the
outer boundary and positive near the inner. In addition, equatorial
symmetry of the nonaxisymmetric flow is clearly broken, but not in the
case of the axisymmetric (zonal) flow.  A third regime, obtained for
the largest $\Ray_n$, is also explored. In this regime the scaling
$K\sim\Ray_n^{1/8}$ is valid and the balance is between inertial and
viscous forces, with the Coriolis force playing a secondary role. The
zonal flow starts to slowly lose equatorial symmetry, and remains
roughly constant. Our results point to a fourth regime. This would
correspond to fully isotropic turbulence characterised by nearly equal
equatorially symmetric and antisymmetric components of the zonal flow
($\overline{K^s_a}/\overline{K_a}\approx 0.5$).

Large time scales (small corotating frequencies), reminiscent of
those at the onset of burning convection, still prevail at the largest
$\Ray_n$ explored, coexisting with faster time scales. This also seems 
to happen when convection is driven only by differential
heating~\cite{GBNS14}. In this case time scales from the onset of convection
have revealed themselves to quite robust to supercritical changes in $\Ray$ and
$\Ray_n$. This gives us some more confidence in, and motivates further study of, the first estimates of time scales of several
astrophysical scenarios provided in~\cite{GCW18} from the linear
stability analysis of differentially heated convection. 

We have also considered how our results may help our understanding of
the role played by convection in the evolution of type-I X-ray bursts
on accreting neutron stars. We have described the limitations of our
type of modelling, and explored a scenario that suggests interesting
consequences. Following previous studies,~\cite{WHCHPRFSBW04}, we
model the early convective phases of a burst by an increase of
$\Ray_n$. As the burst evolves, convective patterns corresponding to
larger $\Ray_n$ are successively preferred. When convective heat
transport becomes of the same order of nuclear heating and dissipative
cooling, i.e at the boundary between flow regimes II and III, we
assume energy is transferred out to the photosphere and start to
decrease $\Ray_n$ as would occur as helium is exhausted. At this point
the flow patterns have small wave number, are retrograde (propagating
westward) and span a wide colatitudinal region around the
equator. These characteristics, which are also exhibited by buoyant
r-modes~\cite{Hey04} have led to these modes being suggested as a
possible mechanism for burst oscillations. By decreasing $\Ray_n$, our
quite different model also predicts a decrease of corotating frame
frequencies, in agreement with the observed drift of the frequency of
burst oscillations.  Finally, on the basis of standard estimations of
physical properties of an accreting neutron star ocean, some
reasonable predictions of temperature and velocities are provided.

Further research will include a comparison of our results for
temperature dependent internal heating ($\Ray_n$) with a model
considering uniform internal heating ($\Ray_i$), both with an imposed
temperature gradient ($\Ray_e$) at the boundaries. Our preliminary
explorations show that the onset of convection takes in a very similar
fashion when one removes the temperature dependence of the internal
heating.  However one should also investigate the question of whether
the temperature dependence is relevant when convection is fully
developed. Finally, as argued above, the determination of the
transition (at certain critical internal heating) between flow regimes
II and III for stronger externally enforced temperature gradients will
help to shed light on fluid flow regimes more relevant to neutron star
oceans.

\medskip
\section{Acknowledgements}

F. G. was supported by a postdoctoral fellowship of the Alexander von
Humboldt Foundation. The authors acknowledge support from ERC Starting
Grant No. 639217 CSINEUTRONSTAR (PI Watts). This work was sponsored by
NWO Exact and Natural Sciences for the use of supercomputer facilities
with the support of SURF Cooperative, Cartesius pilot project
16320-2018.


%

\end{document}